\documentclass[%
 aip,
 amsmath,amssymb,graphicx,
preprint,%
]{revtex4-1}

\usepackage{amsmath}
\usepackage{xcolor}
\usepackage{float}
\usepackage{makecell}
\usepackage{ragged2e}
\usepackage{calligra}
\usepackage[T1]{fontenc}

\usepackage{dcolumn}

\usepackage[mathlines]{lineno}

\usepackage[utf8]{inputenc}
\usepackage[T1]{fontenc}
\usepackage{graphicx}
\usepackage{subcaption}
\usepackage{booktabs}
\usepackage{multirow}
\usepackage{caption}
\usepackage{bm}
\usepackage{tabularx}
\usepackage{tikz}
\usepackage{overpic}
\PassOptionsToPackage{authoryear}{natbib}
\usepackage[authoryear]{natbib}
\usepackage[colorlinks, linkcolor=blue, anchorcolor=blue, citecolor=blue]{hyperref}

\begin{document} 
\preprint{}

\title[]{Long-horizon prediction of three-dimensional wall-bounded turbulence with CTA-Swin-UNet and resolvent analysis}

\author{Bo Chen}
\thanks{These authors contributed to the work equally.}

\author{Yitong Fan}
\thanks{These authors contributed to the work equally.}

\affiliation{
School of Aeronautics and Astronautics, Shanghai Jiao Tong University, Shanghai 200240, China
}

\author{Jie Yao}

\affiliation{
School of Interdisciplinary Science, Beijing Institute of Technology, Beijing 100081, China
}

\author{Weipeng Li}
\email{liweipeng@sjtu.edu.cn}

\affiliation{
School of Aeronautics and Astronautics, Shanghai Jiao Tong University, Shanghai 200240, China
}

\date{\today}
\begin{abstract}
	Long-horizon prediction of three-dimensional (3D) wall-bounded turbulence with machine-learning methods remains a challenging task, due to the rapid accumulation of autoregressive errors and the substantially computational cost.
    To address these challenges, we present a hybrid machine-learning framework, in which a channel-time-attention Swin-UNet (CTA-Swin-UNet) and a multi-time-scale fusion correction (MTFC) strategy are developed to predict the turbulent flow fields in a wall-parallel plane, with affordable computational cost. Then, 3D flow fields are reconstructed via a resolvent-based spectral linear stochastic estimation (SLSE), rooting from the predicted planar flow.
    Results show that the CTA-Swin-UNet outperforms the baseline models (LSTM, FNO and traditional Swin-UNet) in both single-step prediction and autoregressive rollouts, indicating the effectiveness of introducing the CTA module into the Swin-UNet architecture.
    At the same temporal interval, the CTA-Swin-UNet remains stable for approximately 150 rollout steps, while the baseline models fail within 20 to 50 rollout steps.
    After introducing the MTFC strategy, a longer horizon upto 300 steps is achieved.
    Using the resolvent-based SLSE reconstruction further recovers the 3D flow structures and energy spectral distributions from the predicted planar inputs,  
    which demonstrates that the proposed framework provides an effective and computationally efficient approach for long-horizon autoregressive prediction of 3D wall-bounded turbulence.

\end{abstract}
\pacs{}
\maketitle

%
\nolinenumbers
\section{\label{Intro}Introduction}

For decades, computational fluid dynamics (CFD) has served as the primary tool for resolving wall-bounded turbulence~\cite{Pope2000,SmitsMarusicJimenez2011}, such as direct numerical simulation (DNS) and large-eddy simulation (LES).  Nevertheless, they always take expensive computational cost, especially for high-Reynolds-number flow and complex geometries~\cite{Jimenez2012CascadesIW,Duraisamy2019}. This limitation has spurred growing interest in machine-learning-based surrogate models as an alternative pathway for turbulent flow prediction. Recently, machine learning has been used for turbulence modeling, inflow generation~\cite{Beck2021, SoleraRico2024, Brunton2020}, reduced-order modeling, and super-resolution reconstruction, to name a few. 

The use of deep-learning surrogate models holds great promise for long-horizon prediction of turbulent flows \cite{Patil2023}. 
Recurrent neural networks (RNNs), particularly long short-term memory (LSTM) networks, have been used for temporal prediction of turbulent flow fields~\cite{Srinivasan2019,Eivazi2021}.
However, they struggle to capture the multiscale spatiotemporal dynamics of turbulence~\cite{Vlachas2024}, and suffer from rapid error growth during long autoregressive rollouts~\cite{Patil2023}.
Accordingly, integrating physical constraints with LSTM-based methods has been proposed to enhance autoregressive prediction of turbulent flows.
Wang \emph{et al.}~\cite{Wang2020TFNet} proposed TF-Net, which incorporates a divergence-free penalty into a ConvLSTM architecture to enforce mass conservation.
For two-dimensional (2D) Rayleigh-Bénard convection, this constraint reduces long-horizon errors compared with data-driven baselines such as ResNet, U-Net, and ConvLSTM.
Ren \emph{et al.}~\cite{Ren2022PhyCRNet} developed PhyCRNet, where a ConvLSTM encoder-decoder is trained with a discretized Partial Differential Equation (PDE) residual loss.
This design improves long-horizon extrapolation for nonlinear PDE systems, including 2D Burgers and reaction-diffusion equations.

Beyond RNN-based models, attention-based architectures have been explored to suppress the error accumulation during autoregressive rollouts, owing to their abilities in capturing long-range correlations.
Patil \emph{et al.}~\cite{Patil2022AutoRegTrans} proposed an autoregressive transformer for 2D homogeneous isotropic turbulence.
Compared with Fourier neural operator (FNO), their model provides more accurate short-term predictions and shows improved stability during autoregressive rollout.
Li \emph{et al.}~\cite{TNO2024} proposed a transformer neural operator and tested on homogeneous isotropic turbulence and a free-shear turbulent mixing layer. It outperformed FNO and LES with a dynamic mixed model.
Yang \emph{et al.}~\cite{Yang2026IFactFormer} developed a modified implicit factorized transformer for the prediction of turbulent channel flow.
The method reduces short-term prediction errors and maintains more stable long-term statistics.
 
Despite these advances, applying transformer-based architectures to predict full 3D wall-bounded turbulence remains challenging.
Standard global self-attention incurs $\mathcal{O}(n^2)$ computational and memory complexity in the number of spatial tokens, making it prohibitively expensive for high-dimensional turbulence prediction~\cite{Vaswani2017Attention,Dosovitskiy2021ViT}. 
One common strategy is to first map high-dimensional flow fields into a compact low-dimensional representation and then learn its temporal evolution through transformer-based architectures. 
A classical method is based on linear modal decomposition, typically using Proper Orthogonal Decomposition (POD), which represents the flow field using a truncated set of energetic modes. 
Transformer networks are used to predict the temporal evolution of the POD modal coefficients~\cite{Wu2022TransformerROM}. 
Another method is to construct nonlinear reduced representations with autoencoder-based models, where compact latent variables are extracted from the flow fields and subsequently temporal dynamics are then predicted by transformer-based models~\cite{Hemmasian2023ROMTransformer,SoleraRico2024BetaVAETransformer}.

Shifted-window attention in transformer architectures, as introduced in the Swin Transformer~\cite{Liu2021Swin}, also provides an effective way to reduce the computational and memory burden of high-dimensional turbulent flow prediction by restricting self-attention to local windows while enabling cross-window information exchange, which reduces the attention complexity from $\mathcal{O}(n^2)$ to approximately $\mathcal{O}(n)$ with respect to the number of spatial tokens. 
Recent studies have begun to explore Swin-based architectures in turbulence-related tasks. 
Zhang \emph{et al.}~\cite{Zhang2023SwinCompress} developed a Swin-Transformer-based framework for efficient compression of turbulent flow data, showing that hierarchical windowed attention can preserve dominant flow information at reduced storage cost. 
Wang \emph{et al.}~\cite{MSST2024SwinSR} proposed a multi-stage Swin-Transformer network for super-resolution reconstruction of turbulent flows, using progressive feature refinement to recover high-resolution flow structures from coarse inputs. 
Furthermore, Liu \emph{et al.}~\cite{Liu2025MHASTRTurb}  introduced a multi-scale hybrid-attention Swin-Transformer model to enhance turbulence super-resolution, demonstrating improved reconstruction of fine-scale structures through combined local and multi-scale attention mechanisms. 
Nevertheless, these studies remain focused mainly on compression or super-resolution reconstruction, and few studies have focused on the use of Swin-based architectures for Long-horizon prediction of three-dimensional wall-bounded turbulence.

In the present work, a Swin-UNet based network coupled with channel-time-attention is introduced.
However, applying this architecture directly to the full 3D flow field still requires substantial computational and memory resources. 
To further reduce the learning cost, we train the model on a 2D wall-parallel plane and perform autoregressive rollouts on this plane, while recovering the corresponding 3D flow fields through a resolvent-based spectral linear stochastic estimation (SLSE) formulation.
Conventional LSE identifies the reconstruction operator from two-point cross-spectral statistics, which usually require fully resolved ensemble data~\cite{Mathis2009,Baars2016}.
To avoid this dependence, LSE can be reformulated in spectral space and combined with resolvent analysis, so that the reconstruction operator is derived from the mean flow and fluid properties through a resolvent-based linear model~\cite{Rolandi2024Invitation}.
This formulation provides a physics-informed mapping from wall-parallel velocity fluctuations to three-dimensional energetic structures~\cite{Sharma2013}.
For instance, Illingworth \emph{et al.}~\cite{Illingworth2018} used a linearized Navier--Stokes model with eddy viscosity to estimate large-scale streamwise structures in turbulent channel flow from single-plane velocity measurements.
Towne \emph{et al.}~\cite{Towne2020ResEst} further developed a resolvent-based estimation framework to recover space--time flow statistics from limited flow measurements.

Building on this idea, we propose a hybrid framework for long-horizon autoregressive prediction of 3D wall-bounded turbulence.
First, a channel-time-attention Swin-UNet (CTA-Swin-UNet) is developed to predict the temporal evolution on a selected wall-parallel plane. By incorporating a CTA module into the Swin-UNet architecture, the CTA-Swin-UNet improves the use of multi-frame inputs and three-component velocity information.
To improve the stability of long autoregressive rollouts, a multi-time-scale fusion correction (MTFC) strategy is introduced to reduce accumulated errors by combining predictions evolved at different temporal scales.
The corrected long-horizon planar fields are further lifted to the   3D fields through a resolvent-based spectral linear stochastic estimation (SLSE) formulation.
The resulting framework integrates data-driven planar forecasting, multi-time-scale error correction, and physics-informed spectral reconstruction, providing a computationally tractable route for temporally stable prediction of high-dimensional 3D turbulent flows.

The remainder of this paper is organized as follows. Section~\ref{sec:methods} describes the dataset, network architecture, and MTFC strategy. Section~\ref{sec:results} presents and discusses the results, encompassing single-step prediction accuracy, long-horizon autoregressive stability, the effectiveness of MTFC, and 3D flow field reconstruction. Section~\ref{sec:conclusions} summarizes the main findings and outlines directions for future work.

\section{\label{sec:methods}Methods}

\begin{figure}[htbp]
	\centering
	\includegraphics[width=\textwidth,trim=50 110 30 150,clip]{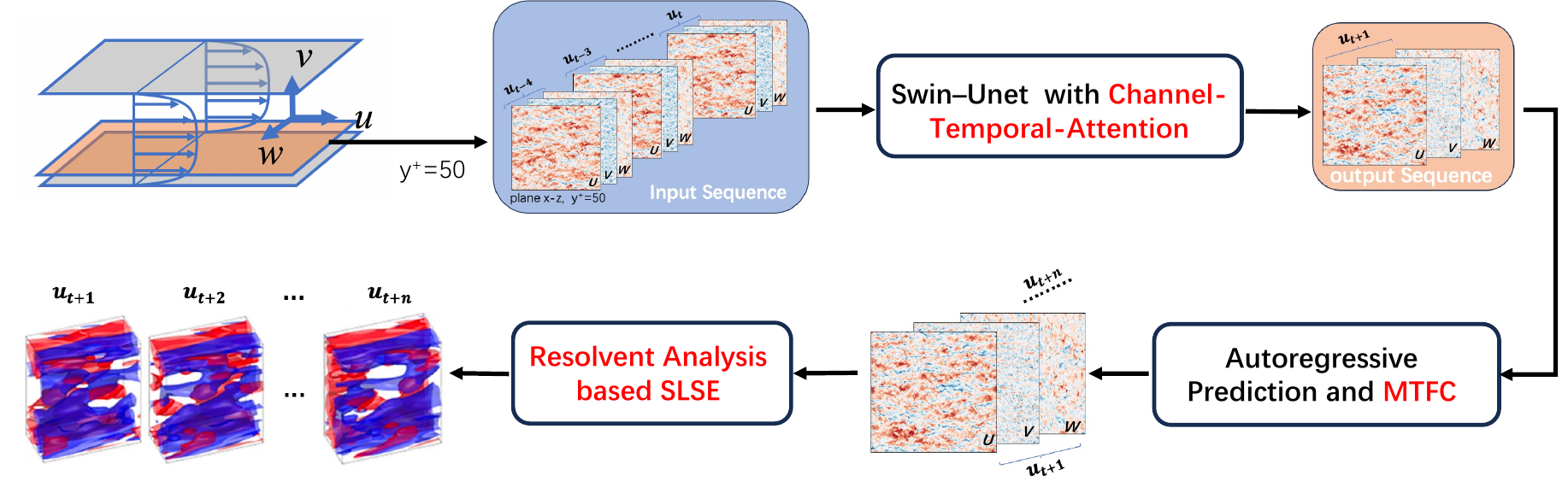}
	\caption{Schematic of the proposed 3D wall-bounded turbulence reconstruction framework.}
	\label{fig:frame}
\end{figure}

Figure~\ref{fig:frame} shows the structure of the proposed framework.
A channel-time-attention Swin-UNet (CTA-Swin-UNet) is developed to predict the temporal evolution, taking the wall-parallel plane fields as the input. The CTA-Swin-UNet is trained as a one-step predictor and then deployed autoregressively for long-horizon rollout. In order to mitigate error accumulation during long-horizon rollout, a multi-time-scale fusion correction (MTFC) strategy is proposed. The predicted planar time series is then used to reconstruct the 3D flow fields via resolvent-based spectral linear stochastic estimation (SLSE). 
In the following context, the dataset setup is described in Section~\ref{sec:dataset}. 
The CTA-Swin-UNet architecture is then introduced in Section~\ref{sec:architecture}. 
The MTFC strategy is presented in Section~\ref{sec:autoregressive}. 
Finally, the SLSE reconstruction procedure is detailed in Section~\ref{sec:3drecon}.

\subsection{\label{sec:dataset}Dataset Setup}

The training and testing of the proposed framework are based on DNS data of 3D incompressible turbulent channel flow at $Re_\tau \approx 550$. The numerical methodology and validation of the DNS simulation are described in Ref.~\onlinecite{Yao2025OppositionControlHighRe}. The computational domain has a size of $L_x \times L_y \times L_z = 4\pi \times 2 \times 2\pi$ in the streamwise, wall-normal, and spanwise directions, and is discretized on a grid of $N_x \times N_y \times N_z = 768 \times 384 \times 512$. The dataset comprises 10,000 consecutive snapshots of the velocity fields, storing the streamwise, wall-normal, and spanwise velocity fluctuations $(u,v,w)$ at each snapshot.
The temporal interval between the neighboring snapshots is approximately $0.2\%$ of a large-eddy turnover time. This time resolution is sufficiently fine to capture the rapid near-wall dynamics.

\subsection{\label{sec:architecture}The CTA-Swin-UNet architecture and Training Configuration}

\begin{figure}[htbp]
	\centering
	\includegraphics[width=\textwidth,trim=120 10 120 0,clip]{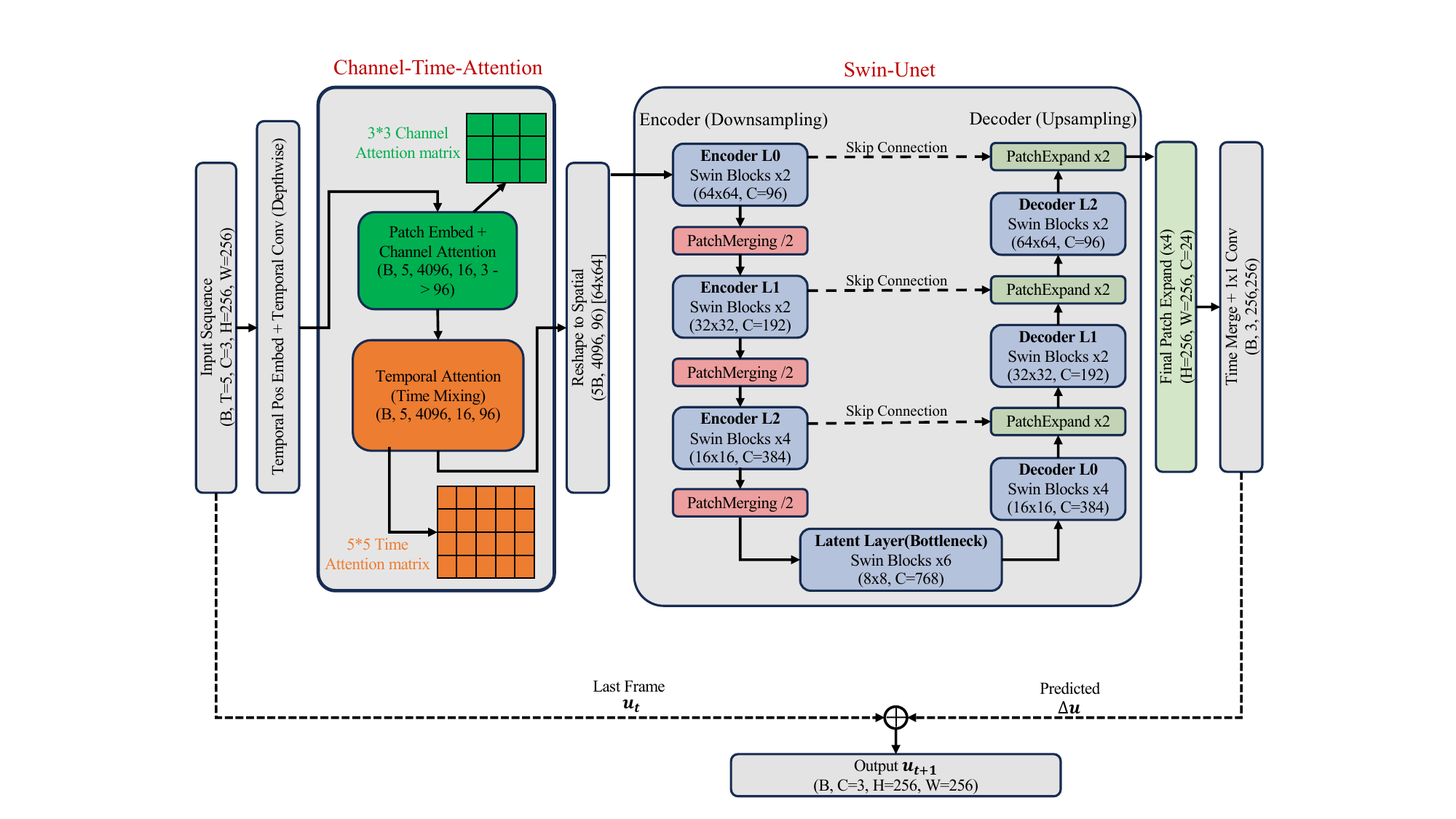}
	\caption{The proposed CTA-Swin-UNet architecture. }
	\label{network_all}
\end{figure}

The architecture of the proposed CTA-Swin-UNet is displayed in Fig.~\ref{network_all}. The model takes five wall-parallel ($y^+=50$) velocity snapshots $\{\mathbf{u}_{t-4},\ldots,\mathbf{u}_{t}\}$ as the input and predicts $\mathbf{u}_{t+1}$ at the next time instant on the same plane. On this wall-parallel plane, stride slicing is applied along the streamwise and spanwise directions to yield a uniform $256\times256$ resolution. Each snapshot $\mathbf{u}_{t}\in\mathbb{R}^{C\times H\times W}$ contains $C=3$ velocity components on a spatial grid of $H\times W=256\times256$. Stacked across a batch size $B$, the input forms a tensor of shape $B\times T\times C\times H\times W$ with $T=5$. The output is formulated in residual form, so that the model predicts the increment with respect to the last input frame rather than the 3D full future fields directly.

As shown in the leftmost two blocks of Fig.~\ref{network_all}, a learnable temporal positional embedding is first used to inject the ordering of the input snapshots. A depthwise $3\times 3$ convolution is applied independently to each velocity component at each time step, providing a learnable spatial filter to reweight neighboring grid points before patchification.
The resulting feature is then fed to the Channel-Time-Attention module shown in Fig.~\ref{network_all}, which consists of a channel-attention sub-block and a temporal-attention sub-block. This ordering first forms component-coupled spatial tokens at each time level and then models their temporal evolution across the input window.

In the channel-attention sub-block, which fuses patchification with cross-component attention, the feature is first partitioned into non-overlapping spatial patches of size $p_h\times p_w$, yielding $\mathbf{P}\in\mathbb{R}^{B\times T\times N\times (p_h p_w)\times C}$, where $N=(H/p_h)(W/p_w)$ is the number of spatial patches. A per-channel linear layer then projects each patch from $p_h p_w$ to $d_c$ dimensions. The result is rearranged to $\mathbf{F}_{c}\in\mathbb{R}^{(BTN)\times C\times d_{c}}$, so that self-attention can be performed across the three velocity components at each spatial patch and time index. The query, key, and value tensors are obtained by linear projections:
\[
\mathbf{Q}_{c}=\mathbf{F}_{c}\mathbf{W}_{c}^{Q},\qquad
\mathbf{K}_{c}=\mathbf{F}_{c}\mathbf{W}_{c}^{K},\qquad
\mathbf{V}_{c}=\mathbf{F}_{c}\mathbf{W}_{c}^{V},
\]
and the channel-attention output is computed as
\[
\mathbf{F}_{c}^{\prime}
=
\mathrm{Softmax}
\left(
\frac{\mathbf{Q}_{c}\mathbf{K}_{c}^{\top}}{\sqrt{d_{c}}}
\right)\mathbf{V}_{c},
\]
where $d_c$ is the scaling dimension. This operation is intended to model the correlations among the streamwise, wall-normal, and spanwise velocity components. The result is then flattened across the channel dimension and linearly projected from $C\cdot d_c$ to the embedding dimension $D$, producing the token representation $\mathbf{F}^{\prime}\in\mathbb{R}^{B\times T\times N\times D}$.

Temporal self-attention is then applied to these channel-refined tokens. They are rearranged to $\mathbf{F}_{t}\in\mathbb{R}^{(BN)\times T\times D}$, so that attention is performed across the input snapshots at each spatial patch. The corresponding query, key, and value tensors are
\[
\mathbf{Q}_{t}=\mathbf{F}_{t}\mathbf{W}_{t}^{Q},\qquad
\mathbf{K}_{t}=\mathbf{F}_{t}\mathbf{W}_{t}^{K},\qquad
\mathbf{V}_{t}=\mathbf{F}_{t}\mathbf{W}_{t}^{V},
\]
and the temporal-attention output is
\[
\mathbf{F}_{t}^{\prime}
=
\mathrm{Softmax}
\left(
\frac{\mathbf{Q}_{t}\mathbf{K}_{t}^{\top}}{\sqrt{d_{t}}}
\right)\mathbf{V}_{t},
\]
where $d_t$ is the corresponding scaling dimension. After reshaping, the final CTA output is obtained as
\[
\mathbf{F}^{\prime\prime}\in\mathbb{R}^{B\times T\times N\times D}.
\]

The tokens processed by CTA are then passed to the Swin-UNet block shown on the right of Fig.~\ref{network_all}. The block consists of a multi-stage encoder, a bottleneck layer, and a multi-stage decoder. In the encoder, patch-merging operations progressively reduce the spatial resolution and increase the channel dimension, yielding hierarchical multi-scale representations. The bottleneck further processes the latent feature at the coarsest scale, and the decoder restores the spatial resolution through patch-expanding layers. Skip connections between encoder and decoder stages of the same scale retain local spatial information. After the final patch expansion, the temporal dimension is merged and a $1\times1$ convolution head produces the predicted velocity increment $\Delta\hat{\mathbf{u}}$. Further architectural details of the Swin-UNet backbone can be found in Ref.~\onlinecite{Cao2022SwinUNet}. The one-step prediction is therefore expressed as
\[
	\hat{\mathbf{u}}_{t+1}=\mathbf{u}_{t}+\Delta\hat{\mathbf{u}}.
\]

For training, each of the three velocity components is normalized
independently using its own mean and standard deviation, computed
on the training split only.
Supervised samples are then constructed by a sliding-window strategy with a stride of one snapshot, producing 9,995 input-target pairs from the 10,000 available snapshots. These pairs are partitioned chronologically into training, validation, and test subsets with
a ratio of 70\%/15\%/15\%, used for parameter optimization,
hyperparameter tuning with early stopping, and evaluation of
single-step and long-horizon performance, respectively. 
The model is trained by minimizing the mean-squared error between the predicted and DNS target fields in the normalized variable space,
\[
\mathcal{L}
=
\frac{1}{CHW}
\sum_{c=1}^{C}
\sum_{h=1}^{H}
\sum_{w=1}^{W}
\left(
\hat{\mathbf{u}}_{c,h,w}^{\,t+1}
-
u_{c,h,w}^{\,t+1}
\right)^2 ,
\]
where $C$ is the number of velocity components and $H \times W$ is the spatial resolution of the wall-parallel plane.
All three velocity components are weighted equally in the loss.
The network is optimized with AdamW, using a warmup-cosine learning-rate
schedule and gradient clipping. The remaining hyperparameters are
summarized in Appendix~\ref{sec:hyperparams}.

\subsection{\label{sec:autoregressive}Multi-Time-Scale Fusion Correction (MTFC)}

After the one-step predictor is trained, long-horizon prediction is performed in an autoregressive manner by recursively feeding previously predicted snapshots back into the input sequence. Let $\mathcal{F}(\cdot)$ denote the trained one-step prediction model. Starting from the initial condition sequence
\begin{equation}
\mathbf{U}_{t-4:t}
=
\left\{
\mathbf{u}_{t-4},
\mathbf{u}_{t-3},
\mathbf{u}_{t-2},
\mathbf{u}_{t-1},
\mathbf{u}_{t}
\right\},
\end{equation}
the first predicted snapshot is obtained as
\begin{equation}
\hat{\mathbf{u}}_{t+1}
=
\mathcal{F}\!\left(\mathbf{U}_{t-4:t}\right),
\end{equation}
and subsequent predictions are generated recursively,
\begin{equation}
\hat{\mathbf{u}}_{t+k}
=
\mathcal{F}\!\left(
\hat{\mathbf{U}}_{t+k-5:t+k-1}
\right),
\qquad k \geq 2.
\end{equation}
Since each predicted snapshot is reused as input for the subsequent prediction, prediction errors propagate from one step to the next and gradually accumulate along the rollout.

Although the CTA mechanism in our architecture is designed to improve the stability of autoregressive rollout, error accumulation is an intrinsic feature of rollout predictions and cannot be addressed by architectural choices alone. To explicitly handle this accumulation at the inference process, we propose a Multi-Time-Scale Fusion Correction (MTFC) strategy that couples two networks trained at different temporal resolutions. The first network, the small-scale model (S-SM) denoted $\mathcal{F}_s(\cdot)$, is the one-step predictor defined above, and advances the flow one snapshot at a time. The second network, the large-scale model (L-SM) denoted $\mathcal{F}_l(\cdot)$, shares the same CTA-Swin-UNet architecture as the S-SM but is trained on a coarser, temporally subsampled sequence drawn from the same DNS data, so that each $\mathcal{F}_l$ call advances the flow by multiple native steps in one prediction. For a given target instant, $\mathcal{F}_l$ therefore reaches it through far fewer recursive calls than $\mathcal{F}_s$. During rollout, the L-SM prediction is combined with the S-SM prediction at periodic fusion instant through a weighted average. The resulting corrected snapshot then replaces the S-SM output and is fed back into the subsequent rollout.

Formally, let $k_f$ denote a rollout step at which fusion is applied and $t+k_f$ the corresponding target instant. The fused state at this fusion point is defined as
\begin{equation}
\tilde{\mathbf{u}}_{t+k_f}
=
(1-\alpha)\hat{\mathbf{u}}_{t+k_f}^{(s)}
+
\alpha \hat{\mathbf{u}}_{t+k_f}^{(l)},
\label{eq:fusion}
\end{equation}
where $\hat{\mathbf{u}}_{t+k_f}^{(s)}$ and $\hat{\mathbf{u}}_{t+k_f}^{(l)}$ are the S-SM and L-SM predictions at the same target instant, and $\alpha\in[0,1]$ is the fusion weight. Note that both branches evolve autoregressively during rollout, so the L-SM input window at $t+k_f$ is itself assembled from previously predicted states rather than ground-truth snapshots. The fused snapshot $\tilde{\mathbf{u}}_{t+k_f}$ then replaces $\hat{\mathbf{u}}_{t+k_f}^{(s)}$ in the subsequent S-SM rollout. By periodically injecting the L-SM correction, the MTFC scheme is designed to suppress the progressive drift of the S-SM rollout while retaining its short-timescale resolution.

\begin{figure}[htbp]
	\centering
	\includegraphics[width=\textwidth,trim=100 150 170 100,clip]{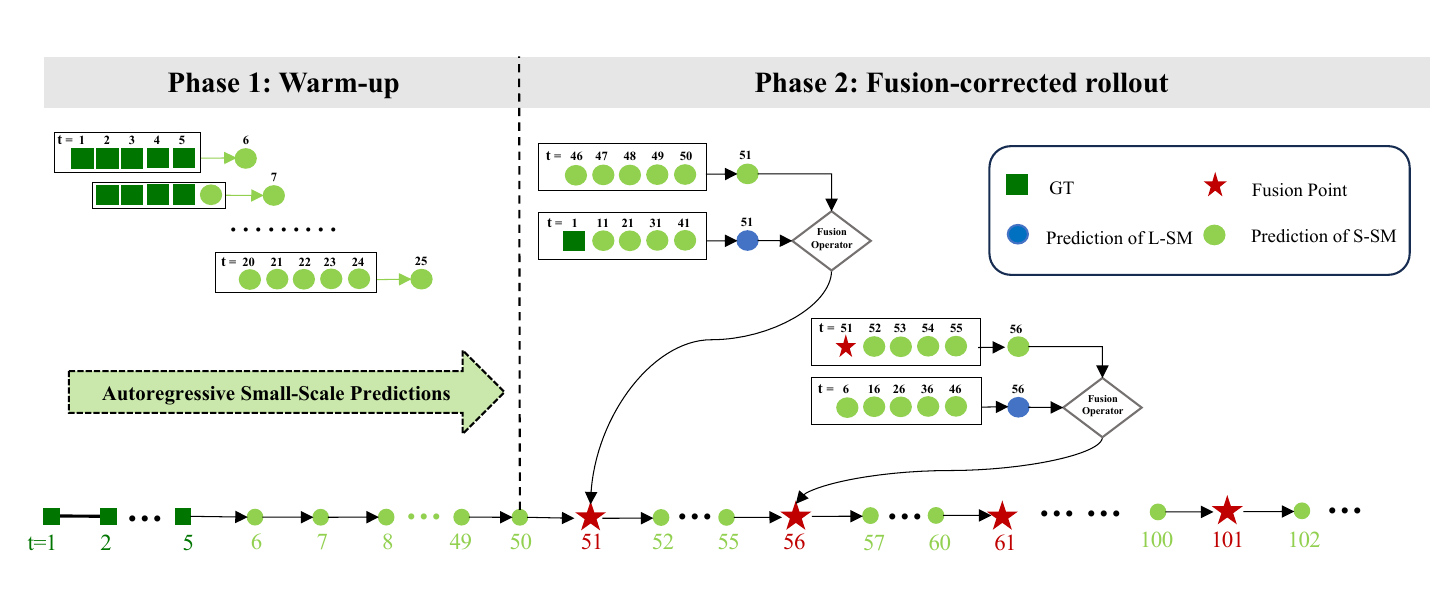}
	\caption{Schematic of the MTFC procedure with fusion interval $fi=5$ steps, first fusion point $ffp=51$ steps, and L-SM stride of ten native steps. Here $fi$ denotes the fusion interval and $ffp$ denotes the first fusion point.\label{fig:mtfc_scheme}}
\end{figure}

To make the fusion procedure concrete, Fig.~\ref{fig:mtfc_scheme} demonstrates a long-horizon rollout with a representative configuration. The L-SM takes an input window $(\mathbf{u}_i, \mathbf{u}_{i+10}, \mathbf{u}_{i+20}, \mathbf{u}_{i+30}, \mathbf{u}_{i+40})$ sampled with a stride of ten native steps, and predicts $\mathbf{u}_{i+50}$. The first fusion point is set to $ffp=51$, the fusion interval to $fi=5$. As shown in the figure, during the first $50$ rollout steps only $\mathcal{F}_s$ is active, advancing the flow one snapshot at a time from the initial window. At the first fusion point $ffp=51$, both models are called for the same target instant. By that point $\mathcal{F}_s$ has executed $45$ autoregressive calls, whereas $\mathcal{F}_l$ has executed only a single forward pass. Combining the two predictions via Eq.~\eqref{eq:fusion} therefore produces a corrected state that draws most of its weight from $\mathcal{F}_l$, which has been advanced through far fewer recursive calls. After this first correction, fusions are repeated every $fi=5$ native steps.

The rationale of the fusion strategy can be understood from two complementary perspectives. The first concerns the autoregressive path length. To reach a given target instant, the fine-timescale model must complete more recursive prediction steps and therefore accumulates greater cumulative drift, whereas the large-timescale model reaches the same instant in fewer steps, typically incurring a smaller total prediction error. Periodic correction from the coarse branch thus effectively re-anchors the fine-timescale trajectory toward a lower-drift reference.
The second concerns the complementary predictive characteristics of the two branches. The fine-timescale model preserves short-term temporal continuity and resolves high-frequency fluctuations, while the large-timescale model better captures slowly varying large-scale structural changes over extended horizons. Their weighted combination improves long-horizon prediction both by suppressing cumulative error growth through periodic correction and by balancing fine-scale fidelity with large-scale structural consistency within a unified prediction framework.

\subsection{3D Flow Fields Reconstruction via Resolvent-Based Spectral Estimation}
\label{sec:3drecon}

The MTFC framework provides long-horizon predictions of all three velocity components $(u,v,w)$ on a single wall-parallel plane at $y^+=50$. While this two-dimensional time series captures the energetically dominant near-wall dynamics, many physical analyses and engineering applications require the 3D volumetric flow fields. Applying CTA-Swin-UNet directly to 3D data would be computationally prohibitive due to the cubic scaling of memory with spatial resolution and the associated attention complexity over a volumetric token set. Instead, we leverage the well-established statistical structure of wall-bounded turbulence, specifically the cross-spectral coherence between different wall-normal positions, to reconstruct the 3D large-scale velocity fields from the single-plane predictions via spectral linear stochastic estimation (SLSE).

\subsubsection*{Spectral linear stochastic estimation}

For the estimation of large-scale fluctuations, a predictive model has been proposed by applying spectral linear stochastic estimation~\cite{Mathis2011,Baars2016}. In the spectral domain, velocity fluctuations at different wall-normal positions are coupled mode-by-mode, for statistically stationary turbulence which is homogeneous in the wall-parallel directions.  
Let $\tilde{u}(\kappa_x, \kappa_z; y, t)$ denote the two-dimensional discrete Fourier transform of the instantaneous streamwise velocity fluctuation $u'$ at a wall-normal position $y$ and time $t$. The cross-spectral density between the estimation position $y_{\mathrm{est}}$ and the reference measurement height $y_{\mathrm{ref}}$ is defined as
\begin{equation}
S_{uu}(\kappa_x, \kappa_z; y_{\mathrm{est}}, y_{\mathrm{ref}})
= \mathbb{E}\!\left[\tilde{u}(\kappa_x, \kappa_z; y_{\mathrm{est}}, t)\,\tilde{u}^\dagger(\kappa_x, \kappa_z; y_{\mathrm{ref}}, t)\right],
\end{equation}
where the superscript $\dagger$ denotes complex conjugation and $\mathbb{E}$ is the expectation operator. The estimation of the fluctuating velocity component can be obtained from a linear expression of the measured velocity $\tilde{u}(\kappa_x, \kappa_z; y_{\mathrm{ref}}, t)$,
\begin{equation}
\tilde{u}(\kappa_x, \kappa_z; y_{\mathrm{est}},t)
= H_L(\kappa_x, \kappa_z;\, y_{\mathrm{est}}, y_{\mathrm{ref}})\;\tilde{u}(\kappa_x, \kappa_z; y_{\mathrm{ref}},t),
\label{eq:slse}
\end{equation}
where
\begin{equation}
H_L(\kappa_x, \kappa_z;\, y_{\mathrm{est}}, y_{\mathrm{ref}})
= \frac{S_{uu}(\kappa_x, \kappa_z; y_{\mathrm{est}}, y_{\mathrm{ref}})}{S_{uu}(\kappa_x, \kappa_z; y_{\mathrm{ref}}, y_{\mathrm{ref}})}.
\label{eq:HL}
\end{equation}

The wavenumber-dependent transfer function is further decomposed by
\begin{equation}
|H_L(\kappa_x, \kappa_z; y_{\mathrm{est}}, y_{\mathrm{ref}})|=\sqrt{\gamma^2(\kappa_x, \kappa_z; y_{\mathrm{est}}, y_{\mathrm{ref}})\frac{|S_{uu}(\kappa_x, \kappa_z; y_{\mathrm{est}}, y_{\mathrm{est}})|}{|S_{uu}(\kappa_x, \kappa_z; y_{\mathrm{ref}}, y_{\mathrm{ref}})|}},
\end{equation}
where $\gamma^2$ is the two-dimensional linear coherence spectrum given by 
\begin{equation}
\gamma^2(\kappa_x, \kappa_z; y_{\mathrm{est}}, y_{\mathrm{ref}})=\frac{|S_{uu}(\kappa_x, \kappa_z; y_{\mathrm{est}}, y_{\mathrm{ref}})|^2}{|S_{uu}(\kappa_x, \kappa_z; y_{\mathrm{est}}, y_{\mathrm{est}})| \cdot|S_{uu}(\kappa_x, \kappa_z; y_{\mathrm{ref}}, y_{\mathrm{ref}})|}\in[0,1].
\end{equation}
$\gamma^2=1$ denotes a linear correlation between the velocity fluctuations at the estimation and measurement plane at the wavenumber combination $(\kappa_x, \kappa_z)$; while $\gamma^2=0$ indicates no linear correlation. 
In this sense, for the estimation of $\tilde{u}_{\mathrm{est}}(\kappa_x, \kappa_z; y_{\mathrm{est}},t)$ from the measurement plane, it is essential to solve the transfer function $H_L(\kappa_x, \kappa_z; y_{\mathrm{est}}, y_{\mathrm{ref}})$.
Normally, computation of $H_L$ from DNS requires ensemble-averaging over various 3D snapshots and is generally unavailable in a deployment setting where only single-plane predictions are accessible. Instead, we derive the cross-spectral densities analytically from the resolvent-based input-output analysis. 

\subsubsection*{Resolvent formulation}

For turbulent channel flows, the linearised non-dimensional Navier--Stokes equations are given by 
\begin{equation}
{\boldsymbol{M}}\frac{\partial \tilde{\boldsymbol{q}}}{\partial t}= {\boldsymbol{A}} \tilde{\boldsymbol{q}} + {\boldsymbol{B}} \tilde{\boldsymbol{f}},
\end{equation}
where $\boldsymbol{q}$ is the state vector including velocity components and pressure, i.e. $\boldsymbol{q}=[u,v,w,p]^\top$, $\boldsymbol{f}$ denotes the nonlinear convection terms in the fluctuating governing equations which is comprised of streamwise ($x$-), wall-normal ($y$-) and spanwise ($z$-) components, i.e. ${\boldsymbol{f}}=[f_x, f_y, f_z]^\top$, and the tilde $\tilde{(\cdot)}$ denotes the Fourier transform in the streamwise and spanwise directions. ${\boldsymbol{M}}$ is a diagonal matrix, with its element being one and zero for the momentum and mass equations, respectively. The coefficient matrices ${\boldsymbol{A}}$ and ${\boldsymbol{B}}$ are given as, respectively, 
\begin{align}
\boldsymbol{A}=\left[\begin{array}{cccc}
-\mathrm{i} k_x \bar{\mathbf{u}} + \nu_T'\frac{d}{ d \boldsymbol{y}}+ \frac{1}{R e} \frac{\nu_T}{\nu} \nabla^2 & -\frac{d \bar{\mathbf{u}}}{d \boldsymbol{y}} + \mathrm{i} k_x \nu_T' \boldsymbol{I} & \mathbf{0} & -\mathrm{i} k_x \boldsymbol{I} \\
\mathbf{0} & -\mathrm{i} k_x \bar{\mathbf{u}} + 2\nu_T'\frac{d}{ d \boldsymbol{y}}
+ \frac{1}{R e}  \frac{\nu_T}{\nu}\nabla^2 & \mathbf{0} & -\frac{d}{d \boldsymbol{y}} \\
\mathbf{0} & \mathrm{i} k_z \nu_T' \boldsymbol{I} & -\mathrm{i} k_x\bar{\mathbf{u}} + \nu_T'\frac{d}{ d \boldsymbol{y}}+ \frac{1}{Re} \frac{\nu_T}{\nu} \nabla^2 & -\mathrm{i} k_z \boldsymbol{I} \\
-\mathrm{i} k_x \boldsymbol{I} & -\frac{d}{d \boldsymbol{y}} & -\mathrm{i} k_z \boldsymbol{I} & \mathbf{0}
\end{array}\right],
\label{eq:L}
\end{align}
and 
\begin{align}
\boldsymbol{B}=\left[\begin{array}{ccc}
\boldsymbol{I} & \boldsymbol{0} & \boldsymbol{0} \\
\boldsymbol{0} & \boldsymbol{I} & \boldsymbol{0} \\
\boldsymbol{0} & \boldsymbol{0} & \boldsymbol{I} \\
\boldsymbol{0} & \boldsymbol{0} & \boldsymbol{0}
\end{array}\right].
\end{align}
where $\nu_T'=\frac{1}{R e} \frac{ \partial \nu_T/\nu}{\partial y}$ and $\nabla^2=d^2 / d \boldsymbol{y}^2-(k_x^2+k_z^2) \boldsymbol{I}$. $Re$ denotes the bulk Reynolds number and $\bar{\mathbf{u}}$ is the mean streamwise velocity.
Herein, a simple eddy viscosity $\nu_t$ is considered, to model part of the nonlinear contribution in the forcing terms. The total eddy viscosity $\nu_T$, which is obtained through the sum of the turbulent eddy-viscosity $\nu_t$ and the kinematic viscosity $\nu$, is modelled by the Cess approximation \cite{Cess1958}, {\it{viz.}}
\begin{equation}
\frac{\nu_T}{\nu}=\frac{1}{2}\left\{1+\frac{\kappa^2 R e_\tau^2}{9}\left(1-\eta^2\right)^2\left(1+2 \eta^2\right)^2\left[1-\exp \left((|\eta|-1) R e_\tau / A\right)\right]^2\right\}^{1 / 2}+\frac{1}{2},
\end{equation} 
where $\eta$ is the wall-normal position ranging from $[-1,1]$ and $Re_\tau$ is the friction Reynolds number. The von Kármán constant is set as $\kappa=0.426$ and the coefficient $A=25.4$, following the previous studies \cite{DelAlamo2006}.

With the forcing $\boldsymbol{f}$ assumed to be stochastic and white in time, i.e. $S_{ff}(t,t')=\delta(t,t')$, where $\delta$ is the Dirac function, the cross-spectral density of the statistically steady-state response can be determined through a related Lyapunov equation,
\begin{equation}
{\boldsymbol{A}} {\boldsymbol{X}} {\boldsymbol{M}} + {\boldsymbol{M}} {\boldsymbol{X}} {\boldsymbol{A}}^{\dagger}+{\boldsymbol{B}} {\boldsymbol{B}}^{\dagger}=\boldsymbol{0},
\label{eq:lyap}
\end{equation}
where ${\boldsymbol{X}}$ is the cross-spectral density of the state vector expressed as ${\boldsymbol{X}}= \mathbb{E}\!\left[\boldsymbol{q} \boldsymbol{q}^{\dagger}\right]$. Normally, the solution to this algebraic Lyapunov equation \eqref{eq:lyap} is not unique due to the non-full-rank matrix ${\boldsymbol{M}}$. In this case, we ignore the response of pressure  fluctuations in the linear operator and instead we treat it as an external stimulus included in the forcing contribution. Consequently, the cross-spectral density of velocity components can be directly solved with a Matlab function `lyap'. With $S_{uu}(\kappa_x, \kappa_z; y_{\mathrm{est}}, y_{\mathrm{ref}})$ and $S_{uu}(\kappa_x, \kappa_z; y_{\mathrm{ref}}, y_{\mathrm{ref}})$ being elements of ${\boldsymbol{X}}$, the transfer function $H_L(\kappa_x, \kappa_z;\, y_{\mathrm{est}}, y_{\mathrm{ref}})$ in \eqref{eq:HL} can be obtained, reconstructing the full boundary layer from the wall-parallel plane predictions with essentially no additional high-fidelity computation.
In this way, the proposed framework recovers the 3D flow fields from the planar predictions without training a network directly on 3D data.

\section{\label{sec:results}Results}

In this section, we first compare the proposed CTA-Swin-UNet against three baselines, namely Swin-UNet without the CTA module, LSTM, and FNO. The comparison is carried out in two stages, single-step prediction across the three velocity components and long-horizon autoregressive rollout examined through error propagation, spectral fidelity, and probe-point time series. We then introduce the MTFC strategy on top of the CTA-Swin-UNet and assess its long-horizon stability over an extended rollout window. Finally, we apply the resolvent-based SLSE to reconstruct the 3D flow fields from the MTFC-predicted reference plane and verify the reconstruction in instantaneous and probe-time-series views. One rollout step corresponds to approximately 5 inner time units $\nu/u_\tau^2$ in the present DNS data.

\subsection{Single-Step Prediction}

We first evaluate the single-step prediction performance of four models: the proposed CTA-Swin-UNet, the plain Swin-UNet (CTA-Swin-UNet without the CTA module), LSTM, and FNO. All model results are shown with their respective best-performing hyperparameter configurations on identical data partitions to ensure fair comparison.

\begin{figure}[htbp]
	\centering
	\includegraphics[width=0.95\textwidth]{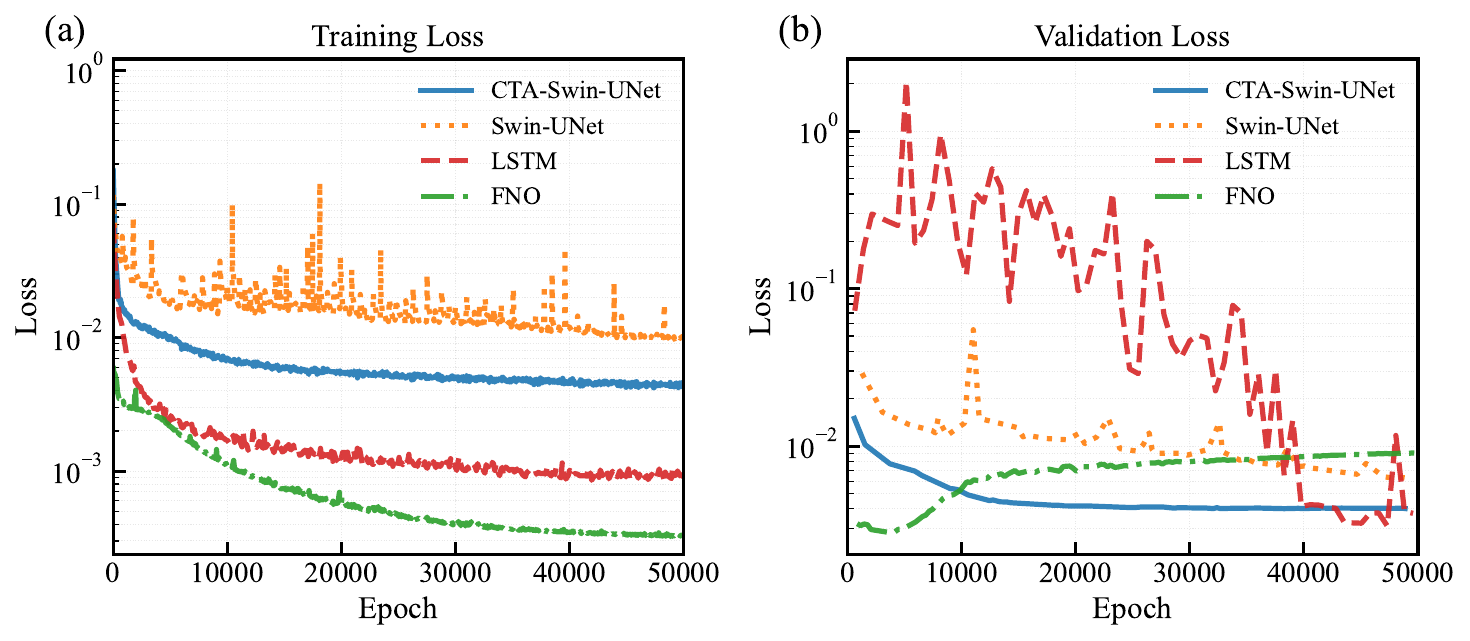}
	\caption{Training and validation loss curves for all four models.}
	\label{fig:training_loss}
\end{figure}

Figure~\ref{fig:training_loss} shows the training and validation loss curves for all four models. 
Although FNO and LSTM achieve substantially lower training errors than CTA-Swin-UNet, their validation errors remain at a comparable level. This indicates a larger gap between training and validation errors for the FNO and LSTM models, suggesting stronger overfitting and weaker generalization. In contrast, CTA-Swin-UNet maintains a smaller gap , which points to better generalization behavior.
Notably, the validation loss of FNO initially decreases but subsequently increases and deviates from the training loss, revealing a persistent overfitting tendency that was not eliminated by standard hyperparameter tuning. 
The plain Swin-UNet gives the largest training and validation errors among the four models.
Table~\ref{tab:single_step_metrics} reports the test-set MSE evaluated over the three velocity components. The comparison is consistent with the validation loss behavior in Fig.~\ref{fig:training_loss}. Among the three components, the streamwise velocity generally exhibits a smaller error than the wall-normal and spanwise components. This behavior may be related to the stronger energetic content and more coherent large-scale organization of the streamwise fluctuations, which make them more predictable from previous wall-parallel snapshots. In contrast, the wall-normal and spanwise components are more strongly associated with smaller-scale turbulent fluctuations, and are therefore more difficult to predict accurately.

\begin{table}[h]
	\centering
	\caption{Quantitative comparison of single-step prediction accuracy across velocity components.}
	\label{tab:single_step_metrics}
	\begin{tabular}{lccc}
		\toprule
		Model & Streamwise ($u$) & Wall-normal ($v$) & Spanwise ($w$) \\
		      & MSE ($\times 10^{-3}$) & MSE ($\times 10^{-3}$) & MSE ($\times 10^{-3}$) \\
		\midrule
		CTA-Swin-UNet & 2.69  & 1.74  & 3.29 \\
		Swin-UNet     & 6.37  & 8.10  & 7.62 \\
		LSTM          & 1.26  & 2.23  & 2.14 \\
		FNO           & 2.32  & 4.25  & 3.98 \\
		\bottomrule
	\end{tabular}
\end{table}

\begin{figure}[htbp]
	\centering
	\includegraphics[width=\textwidth]{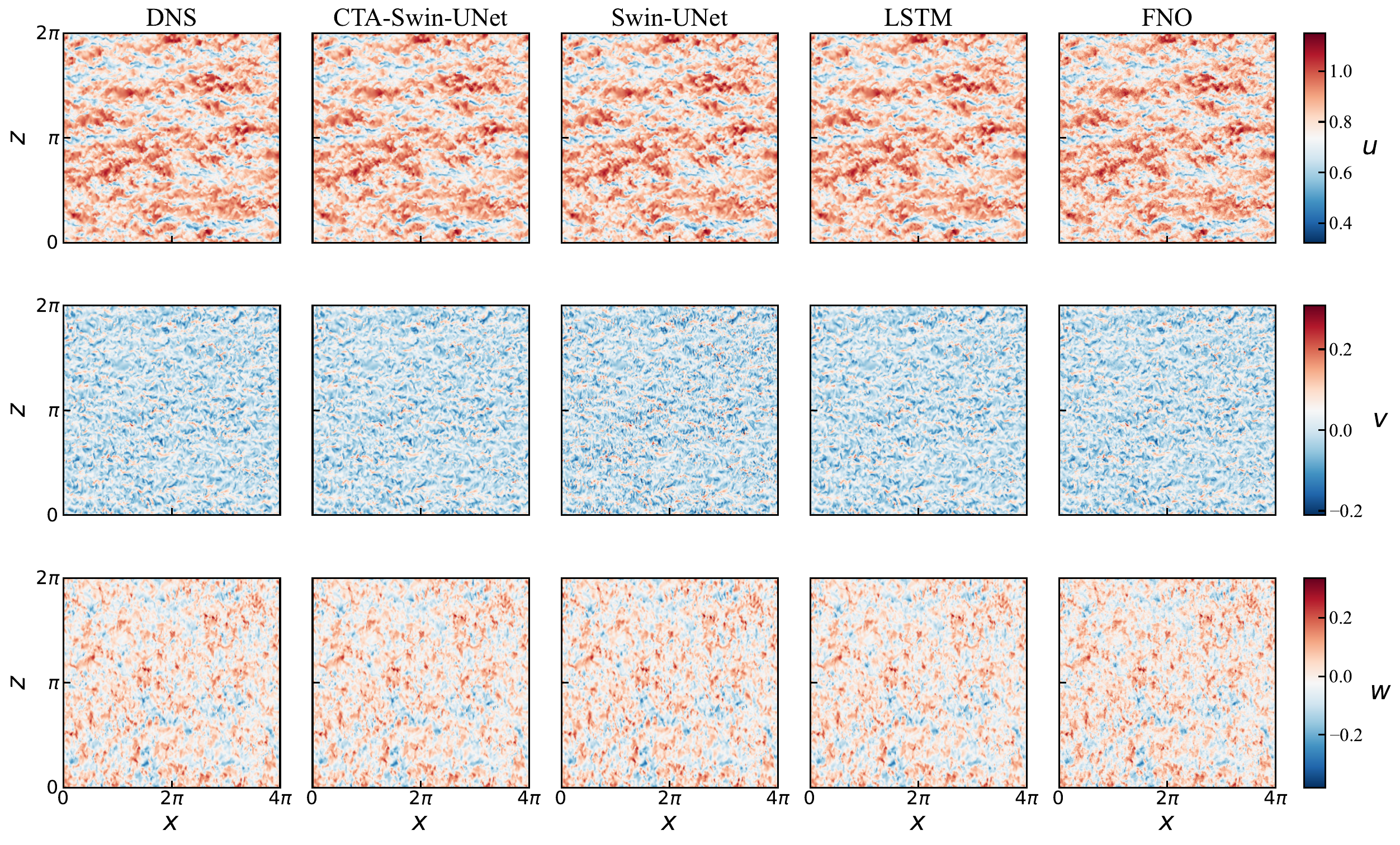}
	\caption{Single-step velocity predictions at a random test timestep. Each row shows one velocity component: streamwise $u$ (top row), wall-normal $v$ (middle row), and spanwise $w$ (bottom row).}
	\label{fig:single_step_viz}
\end{figure}

Despite these quantitative differences, Fig.~\ref{fig:single_step_viz} shows that all four models produce visually accurate instantaneous velocity fields at a randomly selected testing timestep. For the streamwise component $u$ (top row), all models faithfully reproduce the elongated high- and low-speed streaks that dominate the near-wall region, with CTA-Swin-UNet showing the sharpest small-scale features. The wall-normal component $v$ (middle row) and spanwise component $w$ (bottom row) are similarly well reconstructed, with correct velocity magnitudes and spatial coherence across all models. In summary, all four models attain high single-step prediction accuracy, and the visual differences among models remain subtle at the single-step prediction.

\subsection{Autoregressive Rollout}

\begin{figure}[htbp]
	\centering
	\includegraphics[width=\textwidth]{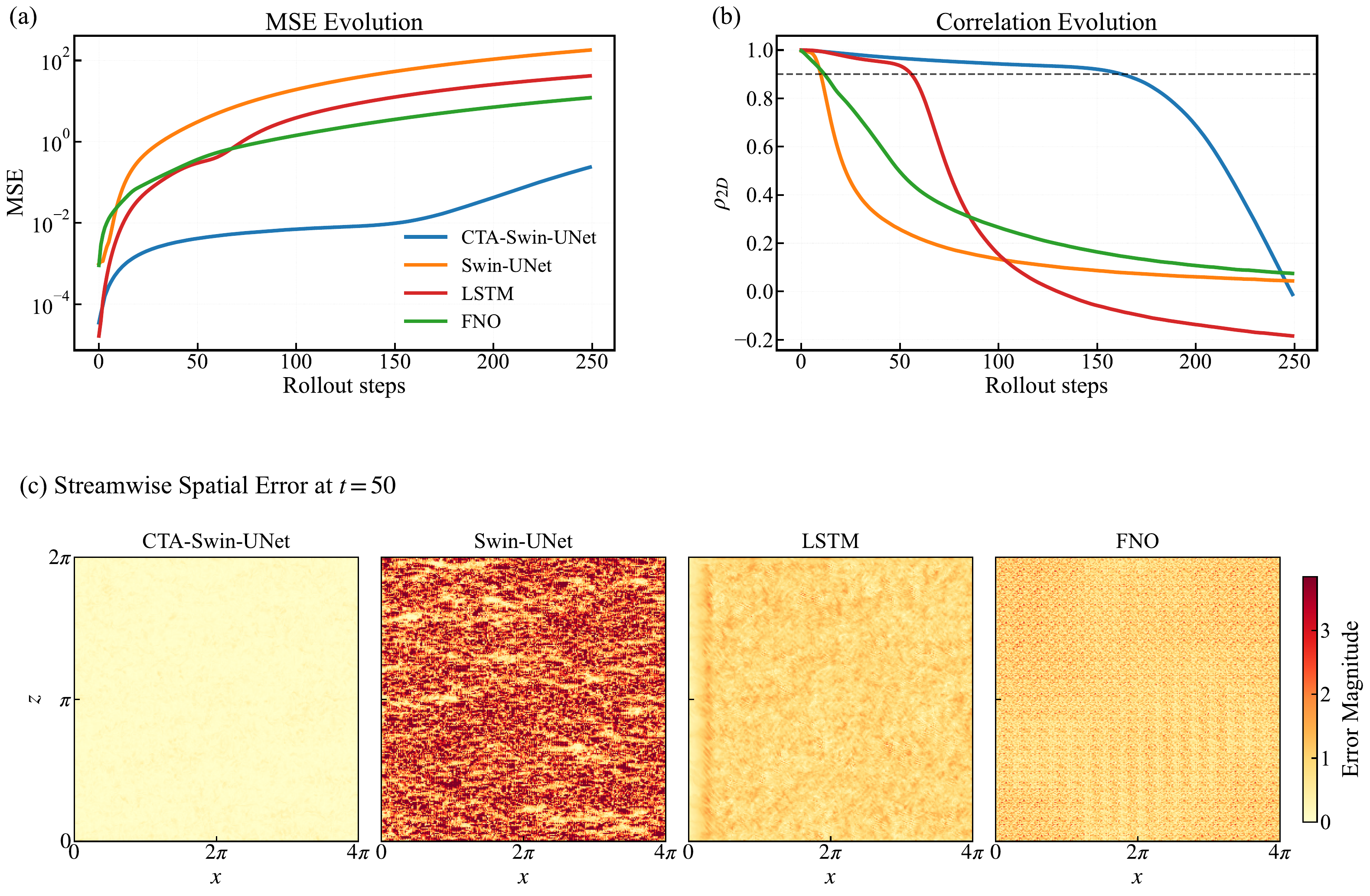}
	\caption{Error propagation during autoregressive rollout. (a) MSE evolution for all four models. (b) Pearson correlation coefficient $\rho$ between each predicted plane and the DNS plane, with the horizontal dashed line marking the stability threshold $\rho=0.9$. (c) Spatial error distribution at $t=50$.}
	\label{fig:error_propagation}
\end{figure}

Temporal prediction of turbulent flows is commonly carried out in an autoregressive manner, with model outputs recursively fed back as inputs for subsequent steps. Long-horizon stability under such rollouts therefore provides a more stringent test than single-step accuracy. In this section, we examine the autoregressive performance of all four models. Figure~\ref{fig:error_propagation} presents the error propagation analysis for all four architectures, with rollouts initialized from the first five consecutive frames of the testing dataset. Panel (a) reveals striking differences in overall error growth rates. The CTA-Swin-UNet maintains remarkably stable predictions for approximately 150 rollout steps. Swin-UNet exhibits the fastest error growth of all four models. Its MSE surpasses those of LSTM and FNO after approximately 20 rollout steps and approaches $10^{2}$ by step 50, indicating complete loss of predictive capability. Both LSTM and FNO models also exhibit significantly faster error growth than CTA-Swin-UNet, with their MSE curves rising sharply beyond 30 rollout steps. Notably, although the LSTM achieves a lower single-step error than CTA-Swin-UNet on this particular initial condition, its autoregressive error accumulates at a substantially faster rate. The dataset-averaged single-step errors in Table~\ref{tab:single_step_metrics} still favour CTA-Swin-UNet, and the rollout result demonstrates that good single-step performance on individual samples does not guarantee stability under long-horizon autoregressive rollout. 

Panel (b) further evaluates the structural similarity between the predicted and DNS results during autoregressive rollout. While the MSE in panel (a) quantifies the absolute amplitude of the prediction error, it is sensitive to the magnitude of different velocity components and does not provide a natural criterion for defining the stable prediction horizon. We therefore introduce the Pearson correlation coefficient\cite{Lippe2023PDERefiner} to measure whether the predicted plane still preserves the spatial organization of the DNS reference.
For each velocity component \(c\in\{u,v,w\}\), the correlation coefficient is computed over the wall-parallel plane as
\begin{equation}
\rho_c(t)
=
\frac{
\sum_{i,j}
\left[
\hat{\mathbf{u}}_c(i,j,t)-\overline{\hat{\mathbf{u}}}_c(t)
\right]
\left[
\mathbf{u}(i,j,t)-\overline{\mathbf{u}}_c(t)
\right]
}{
\sqrt{
\sum_{i,j}
\left[
\hat{\mathbf{u}}_c(i,j,t)-\overline{\hat{\mathbf{u}}}_c(t)
\right]^2
}
\sqrt{
\sum_{i,j}
\left[
\mathbf{u}_c(i,j,t)-\overline{\mathbf{u}}_c(t)
\right]^2
}
},
\label{eq:rho_component}
\end{equation}
where \(\mathbf{u}_c\) and \(\hat{\mathbf{u}}_c\) denote the DNS and predicted velocity components, respectively, \(c\in\{u,v,w\}\) denotes the velocity componentand, and \((i,j)\) denotes the spatial location on the wall-parallel plane. The overbar denotes the spatial average over the plane for each individual component,
\begin{equation}
\overline{\mathbf{u}}_c(t)
=
\frac{1}{N_xN_z}
\sum_{i,j} \mathbf{u}_c(i,j,t).
\end{equation}
The overall planar correlation is then defined as the component-averaged value,
\begin{equation}
\rho(t)
=
\frac{1}{3}
\left[
\rho_u(t)+\rho_v(t)+\rho_w(t)
\right].
\label{eq:rho_avg}
\end{equation}
Since \(\rho(t)\) is bounded in \([-1,1]\), it provides a dimensionless and comparable measure of rollout stability across different models. In this work, the rollout is regarded as stable as long as the component-averaged correlation remains above \(0.9\), which is marked by the horizontal dashed line in panel (b). Under this criterion, the CTA-Swin-UNet maintains a stable rollout for approximately 150 rollout steps, the LSTM for about 50 rollout steps, whereas the plain Swin-UNet and FNO lose stability after about 20 rollout steps.

Panel (c) shows the spatial distribution of errors at $t=50$ steps, where the error refers to the absolute value of the instantaneous streamwise relative error at this time instant. The CTA-Swin-UNet error field remains small and spatially scattered without forming coherent large-scale structures. In contrast, the Swin-UNet error field displays organized streak-like high-error patterns, indicating that accumulated errors have already corrupted the large-scale spatial organization of the flow. The LSTM prediction exhibits extensive regions of systematically elevated error forming large-scale patches, while FNO shows intermediate behavior with errors beginning to organize into coherent structures but not yet fully diverged.

\begin{figure}[htbp]
	\centering
	\includegraphics[width=\textwidth]{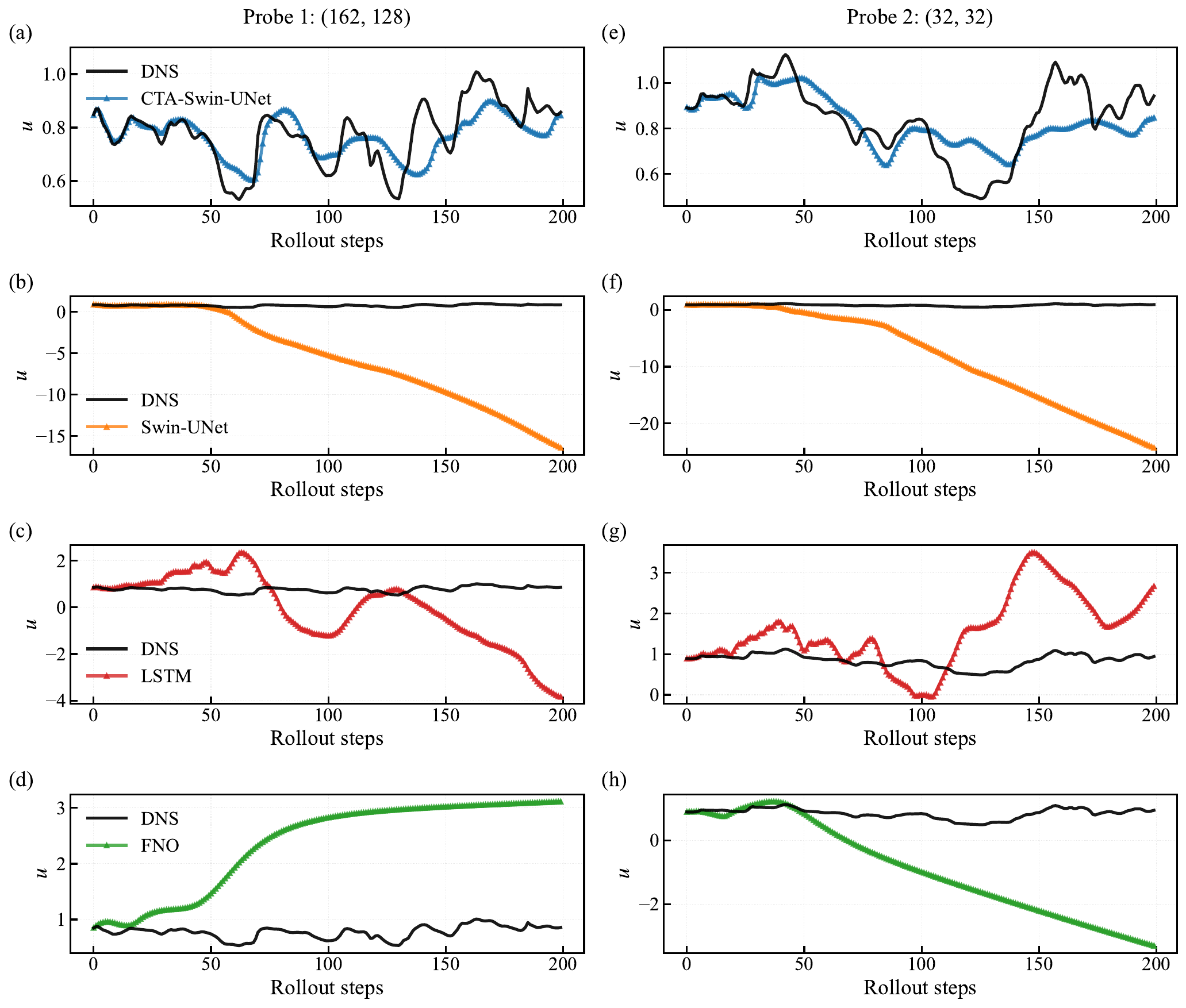}
	\caption{Streamwise velocity time series at two randomly selected probe points. Left column (a-d), probe point 1 at grid indices $(i,j) = (162,128)$; right column (e-h), probe point 2 at grid indices $(i,j) = (32,32)$. Each row corresponds to one model: CTA-Swin-UNet, Swin-UNet, LSTM, and FNO.}
	\label{fig:time_series}
\end{figure}

For a more direct comparison of the autoregressive behaviors, we sample the streamwise velocity at two randomly chosen probe locations and track its temporal evolution. Figure~\ref{fig:time_series} shows the resulting signals. For probe point 1 at grid indices $(i,j) = (162,128)$ (left column), the CTA-Swin-UNet (panel a) maintains near-perfect phase alignment with DNS for approximately 150 rollout steps, accurately tracking both amplitude and frequency of fluctuations. By contrast, Swin-UNet (panel b) shows a distinct failure mode. The velocity amplitude drifts monotonically and reaches unphysical values well outside the DNS fluctuation range by $t \approx 50$ steps. LSTM (panel c) diverges rapidly with oscillatory behavior after approximately $t=20$ steps, while FNO (panel d) shows progressive decorrelation from early steps. Probe point 2 at grid indices $(i,j) = (32,32)$ (right column) confirms the same trend: CTA-Swin-UNet maintains good phase tracking for approximately 150 rollout steps, Swin-UNet again drifts monotonically to unphysical values, and LSTM and FNO diverge within 30 steps.

\begin{figure}[htbp]
	\centering
	\includegraphics[width=\textwidth,trim=0 0 0 50,clip]{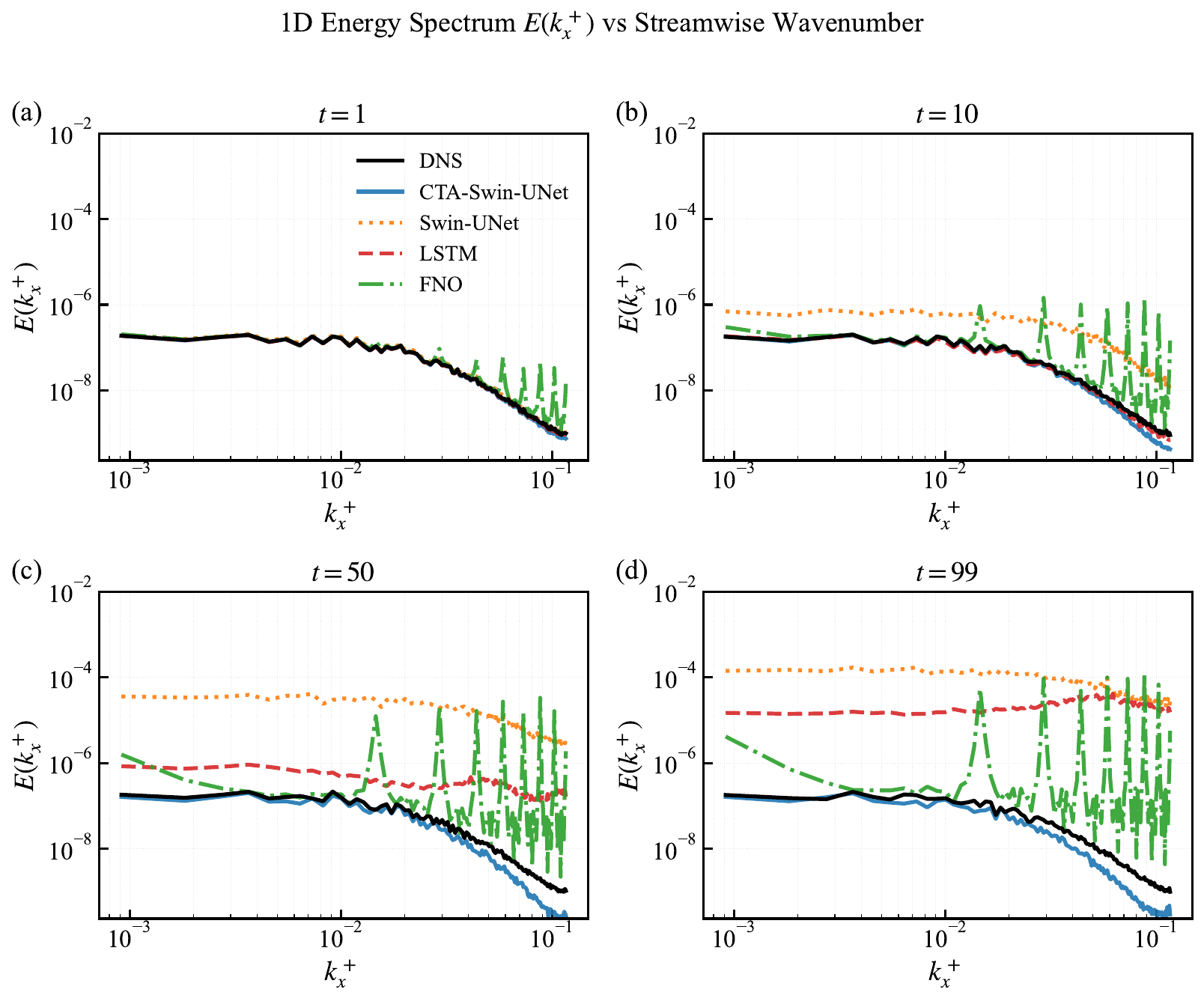}
	\caption{Streamwise one-dimensional energy spectra $E(k_x^+)$ at $t=1,\, 10,\, 50,\, 99$ steps. }
	\label{fig:spectra_kx}
\end{figure}

\begin{figure}[htbp]
	\centering
	\includegraphics[width=\textwidth,trim=0 0 0 50,clip]{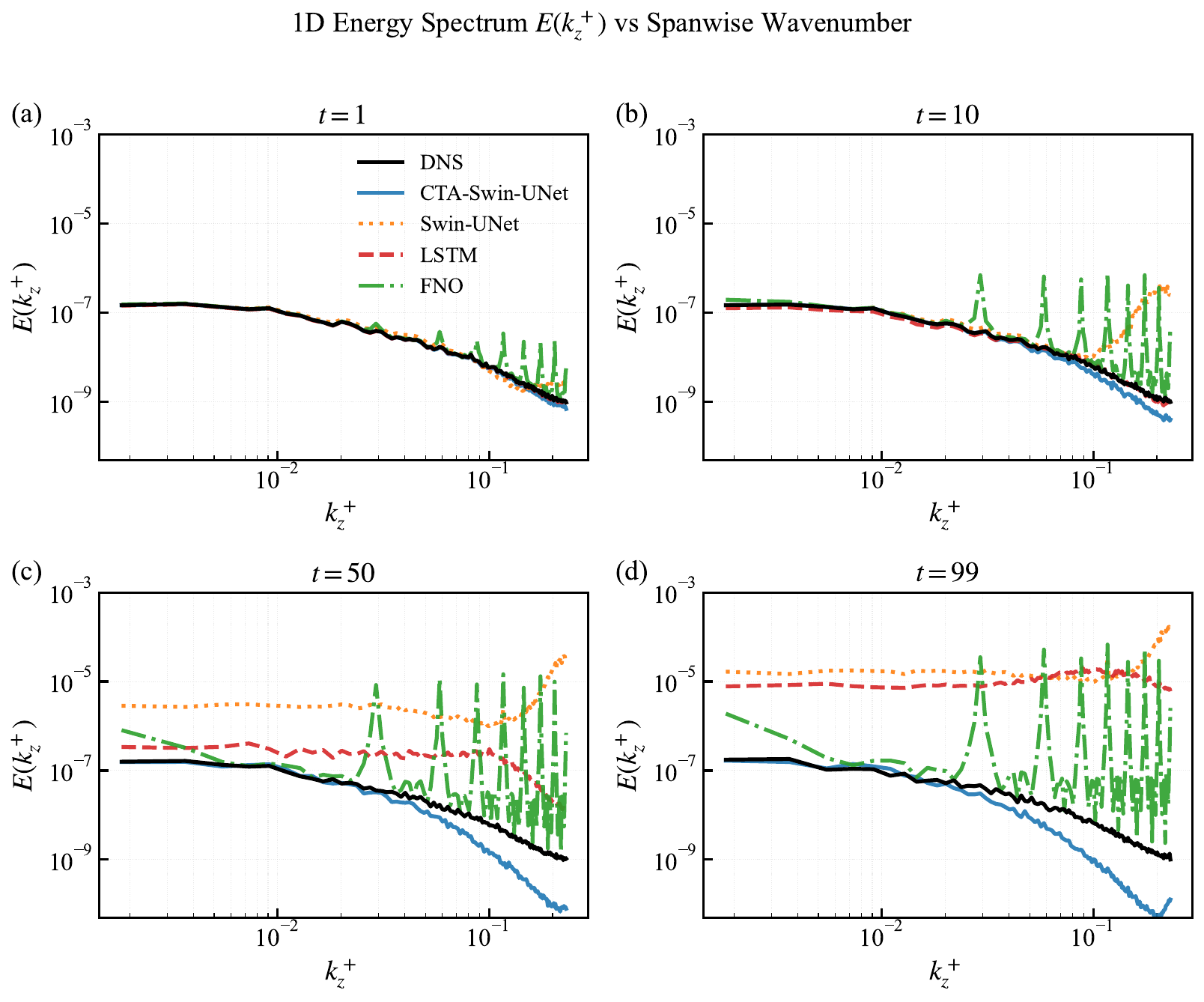}
	\caption{Spanwise one-dimensional energy spectra $E(k_z^+)$ at $t=1,\, 10,\, 50,\, 99$ steps.}
	\label{fig:spectra_kz}
\end{figure}

Beyond the dataset-averaged error propagation in Fig.~\ref{fig:error_propagation}, we further examine the instantaneous spectral fidelity at representative rollout timestepss to assess whether the predictions retain the multiscale energy distribution of DNS.
Figs.~\ref{fig:spectra_kx} and~\ref{fig:spectra_kz} examine the streamwise and spanwise one-dimensional energy spectra $E(k_x^+)$ and $E(k_z^+)$ in wall units at four rollout timestepss. At $t=1$ (panel a), all models except FNO reproduce the DNS spectrum closely; FNO already shows minor fluctuations at high $k_x^+$. By $t=10$ (panel b), Swin-UNet and FNO begin to show noticeable deviations, while CTA-Swin-UNet and LSTM remain well aligned with DNS. At $t=50$ and $t=99$ (panels c and d), Swin-UNet, LSTM, and FNO all exhibit significant spectral deviations; Swin-UNet is most severely affected, with the entire spectrum elevated by two to three orders of magnitude above DNS and flattened into a near-constant level. CTA-Swin-UNet also begins to show deviations at high $k_x^+$ by $t=99$, but overall maintains good agreement with the DNS spectral shape throughout the rollout.
Notably, the earlier deviation at high wavenumbers indicates that small-scale fluctuations lose spectral fidelity before the large-scale energetic motions, which is consistent with the stronger sensitivity of small scales to accumulated autoregressive errors.
The spanwise spectrum $E(k_z^+)$ (Fig.~\ref{fig:spectra_kz}) shows trends consistent with the streamwise analysis. CTA-Swin-UNet maintains the best DNS agreement at all times, while Swin-UNet, LSTM, and FNO show progressively increasing spectral deviations with growing rollout length, and Swin-UNet again exhibits the most severe degradation from $t=50$ onward.

Additional rollouts from different testing initial conditions showed consistent error-growth trends. Taken together, the error propagation, time-series, and spectral analyses consistently demonstrate that CTA-Swin-UNet, benefiting from the channel-time-attention mechanism, substantially outperforms the other three models in autoregressive predictions. For approximately 150 rollout steps, CTA-Swin-UNet maintains reasonably accurate predictions across all three aspects, far exceeding the stable prediction horizon of the baseline models.

\subsection{Long-Horizon Rollout with MTFC}

\begin{figure}[htbp]
	\centering
	\includegraphics[width=0.6\textwidth]{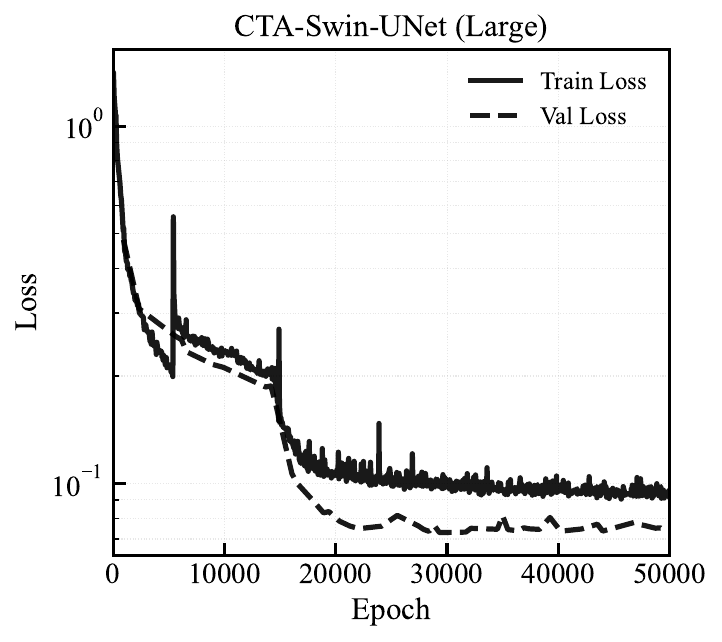}
	\caption{Training and validation loss of the L-SM. }
	\label{fig:training_loss_large}
\end{figure}

The preceding analysis demonstrates that CTA-Swin-UNet achieves superior autoregressive performance, maintaining reasonably accurate predictions for approximately 150 rollout steps. However, as shown in Fig.~\ref{fig:error_propagation}, the error of CTA-Swin-UNet begins to accelerate rapidly beyond this stable window and the model eventually diverges near $t \approx 250$ rollout steps. We therefore introduce the MTFC strategy to further mitigate this error accumulation. As described in Section~\ref{sec:autoregressive}, the core idea is to select an appropriate first fusion point ($ffp$) before the small-scale model (S-SM) diverges, and then apply periodic corrections at a fixed fusion interval ($fi$) using the large-scale model (L-SM) throughout the remainder of the rollout. In this work, an L-SM is trained with a temporal stride 10 times that of the S-SM to investigate the effectiveness of this strategy. The MTFC formulation only requires the L-SM to operate on a temporal scale coarser than that of the S-SM, and varying the specific stride would require retraining a new L-SM, which we leave for future work. Figure~\ref{fig:training_loss_large} shows the training and validation loss of the L-SM, confirming stable convergence on the temporally subsampled sequence without overfitting.

\begin{figure}[htbp]
	\centering
	\includegraphics[width=\textwidth,trim=0 0 0 20,clip]{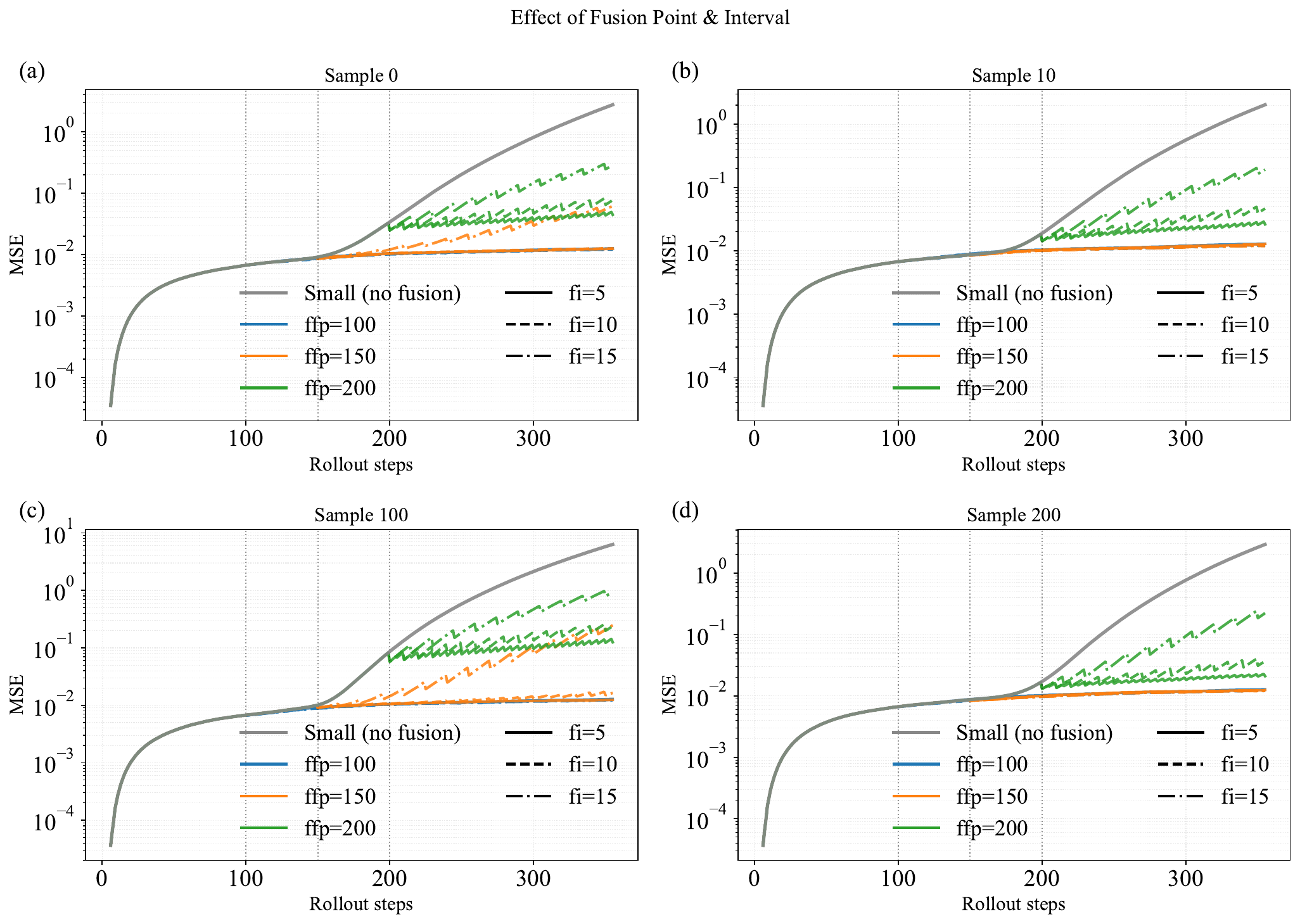}
	\caption{Effect of initial fusion point ($ffp$) and fusion interval ($fi$) on error accumulation during MTFC rollout.}
	\label{fig:fusion_hyperparameter}
\end{figure}

As described in Section~\ref{sec:autoregressive}, the MTFC strategy has three main hyperparameters, the fusion weight $\alpha$ in Eq.~\eqref{eq:fusion}, the first fusion point $ffp$, and the fusion interval $fi$. We first investigate the effect of $ffp$ and $fi$ on the MTFC prediction with the fusion weight fixed at $\alpha=0.8$.
Autoregressive rollouts from four initial input frames sampled from the validation set are performed. 
We evaluate three candidate $ffp$ values $\in \{100, 150, 200\}$ rollout steps. For each $ffp$, we test three fusion intervals $fi \in \{5, 10, 15\}$. The hyperparameter scan is performed jointly over the four samples.
As shown in Fig.~\ref{fig:fusion_hyperparameter},
all four samples show a sharp increase in prediction error around $t = 150$ rollout steps. A smaller fusion interval ($fi = 5$) more effectively suppresses error accumulation. Furthermore, initiating fusions at $ffp$ $= 100$ or $ffp$ $= 150$ yields better long-horizon stability than $ffp$ $= 200$. Since periodic fusion corrections inevitably introduce short-term oscillations, we prefer to delay the first fusion as long as stability allows. 

\begin{figure}[htbp]
	\centering
	\includegraphics[width=0.6\textwidth]{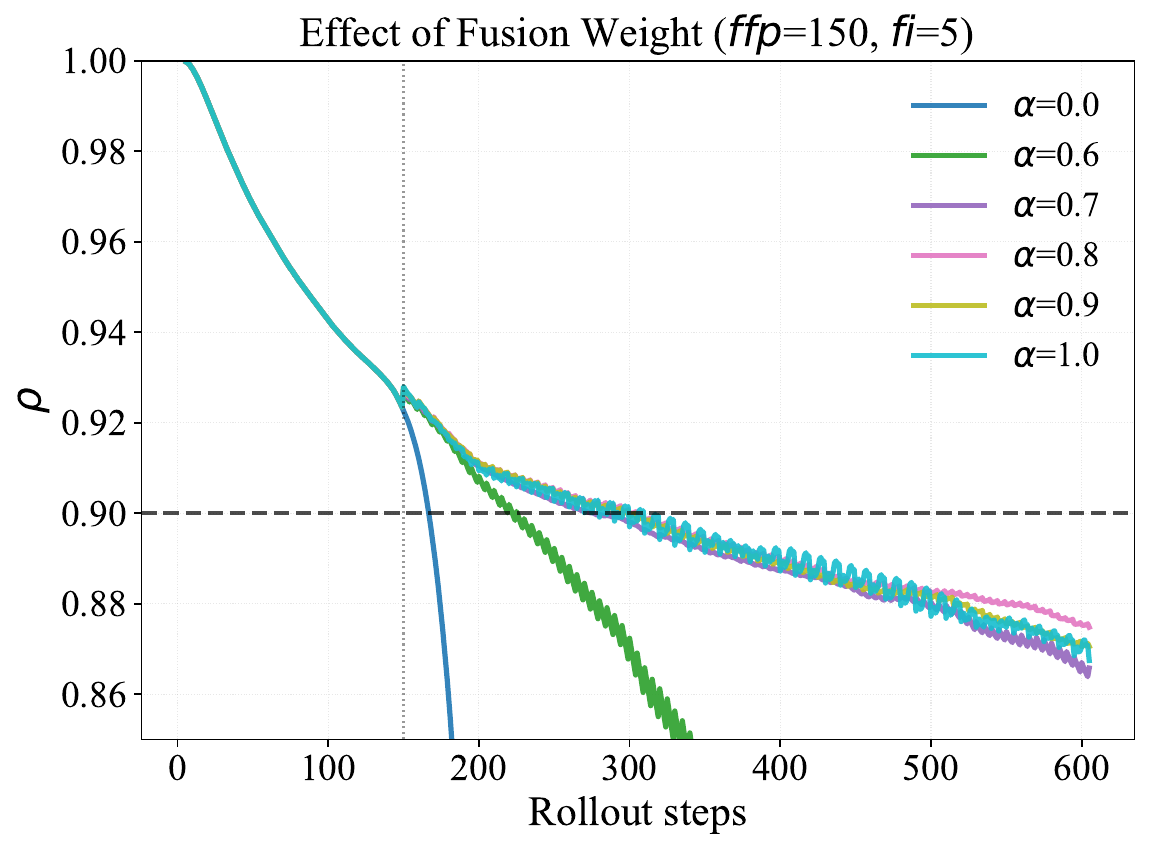}
	\caption{Pearson correlation coefficient $\rho$ versus rollout timesteps for the MTFC framework under $ffp=150$ and $fi=5$, for fusion weights $\alpha \in \{0,\,0.6,\,0.7,\,0.8,\,0.9,\,1\}$. $\alpha=0$ and $\alpha=1$ correspond to using the S-SM or L-SM prediction directly at each fusion point.}
	\label{fig:fusion_weight_corr}
\end{figure}

Building on the above analysis, we further investigate the effect of the fusion weight $\alpha$ on the fused prediction with $ffp=150$ and $fi=5$ fixed. Figure~\ref{fig:fusion_weight_corr} shows the rollout correlation $\rho$ for $\alpha \in \{0,\,0.6,\,0.7,\,0.8,\,0.9,\,1\}$, where $\alpha=0$ and $\alpha=1$ correspond to directly using the S-SM or L-SM prediction at each fusion point, respectively. The fusion strategy becomes effective once $\alpha \geq 0.7$, and over the longer horizon beyond $t=500$ rollout steps $\alpha=0.8$ exhibits the best stability. A possible reason is that when $\alpha$ is raised to $0.9$ or above, the correction injected at each fusion point becomes too strong, amplifying the perturbation seen by the S-SM at its next input window and demanding greater robustness from the model than it can provide.

\begin{figure}[htbp]
	\centering
	\includegraphics[width=0.6\textwidth]{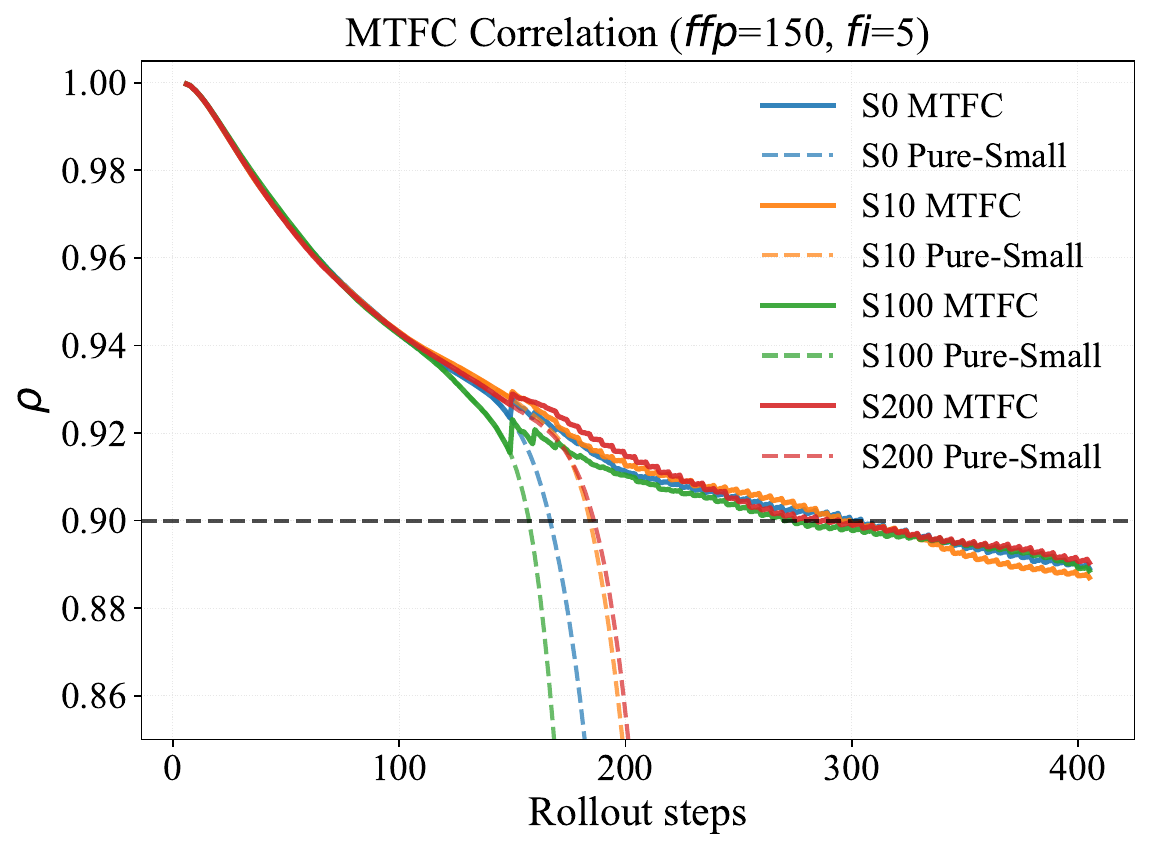}
	\caption{Pearson correlation coefficient $\rho$ versus rollout timesteps for the MTFC (solid lines) and the S-SM alone (dashed lines), evaluated on four samples under $ffp$ $=150$ and $fi=5$. Horizontal dashed lines mark the $\rho=0.9$ stability threshold.}
	\label{fig:fusion_correlation_single}
\end{figure}

Based on these scans, we adopt $\alpha=0.8$, $ffp=150$, and $fi=5$ for all subsequent analyses. Fig.~\ref{fig:fusion_correlation_single} demonstrates the effectiveness of this configuration on all four samples, tracking the rollout correlation $\rho$ for the MTFC framework (solid lines) and the S-SM alone (dashed lines). Across all four initial conditions, the MTFC framework extends stable high-fidelity prediction beyond 300 rollout steps, while the S-SM alone diverges near $t\approx 150$. The detailed analyses below are illustrated on a representative sample (sample 0) for clarity. 

\begin{figure}[htbp]
	\centering
	\includegraphics[width=0.7\textwidth]{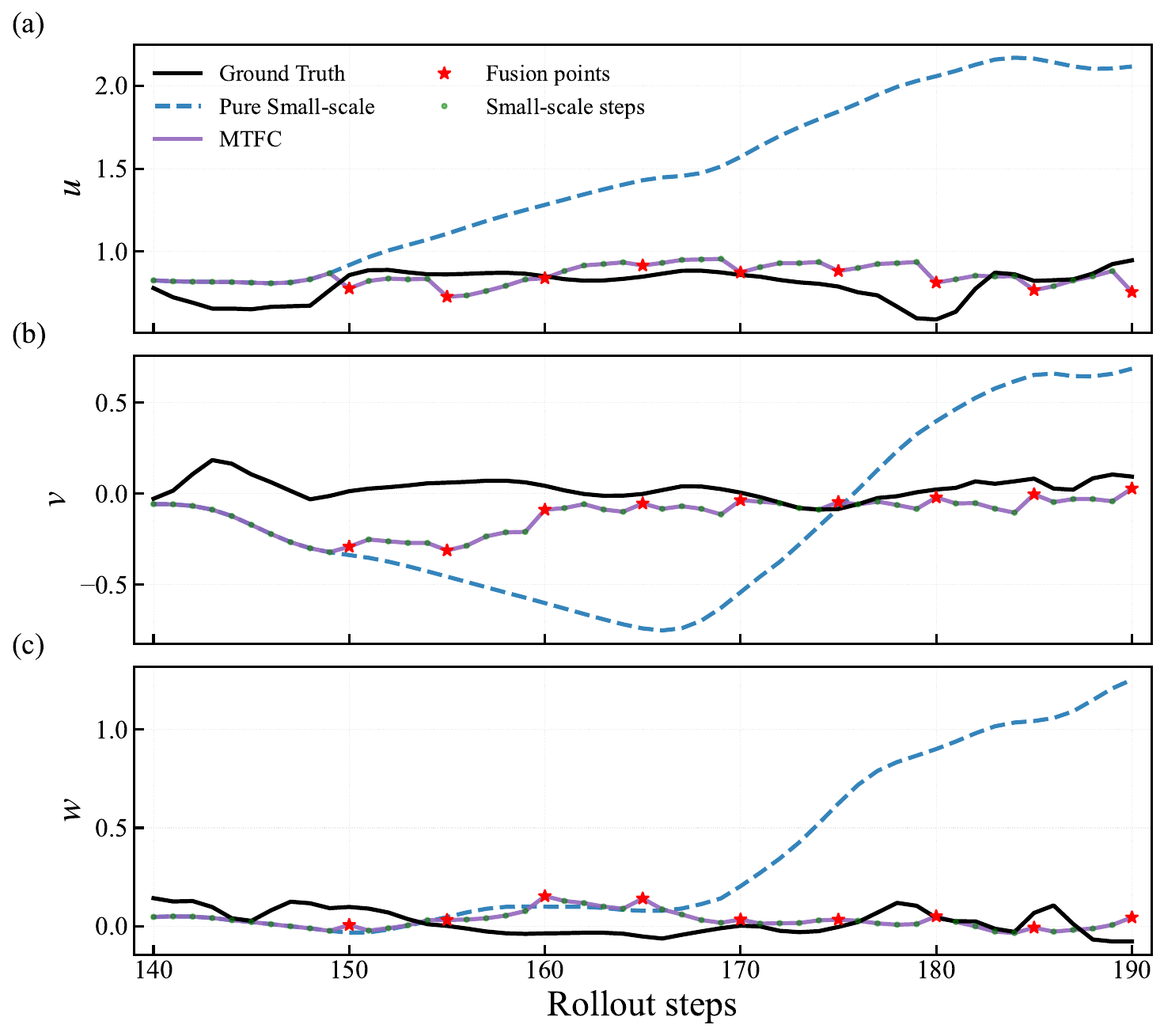}
	\caption{Temporal evolution of the three velocity components at a probe location from $t=140$ to $190$ steps. (a) Streamwise velocity $u$; (b) wall-normal velocity $v$; (c) spanwise velocity $w$. }
	\label{fig:fusion_timeseries}
\end{figure}

The temporal effectiveness of this mechanism is further demonstrated in Figure~\ref{fig:fusion_timeseries}, which tracks the velocity evolution at a probe location over $t=140$ to $190$ steps. In panel (a), the streamwise velocity $u$ from pure S-SM (blue dashed) drifts away from DNS over the rollout, and by $t=190$ the predicted value exceeds the DNS reference by more than twice the local $u_{\rm rms}$. The MTFC (purple solid line) keeps the predicted velocity within the fluctuation range observed in the DNS signal via periodic corrections (red stars), with S-SM steps between corrections (green dots) following the high-frequency dynamics. Wall-normal $v$ (panel b) and spanwise $w$ (panel c) show the same behavior: pure S-SM drifts monotonically, while MTFC remains bounded and physically plausible, confirming that all three velocity components are simultaneously stabilized by the fusion strategy.

\begin{figure}[htbp]
	\centering
	\includegraphics[width=\textwidth]{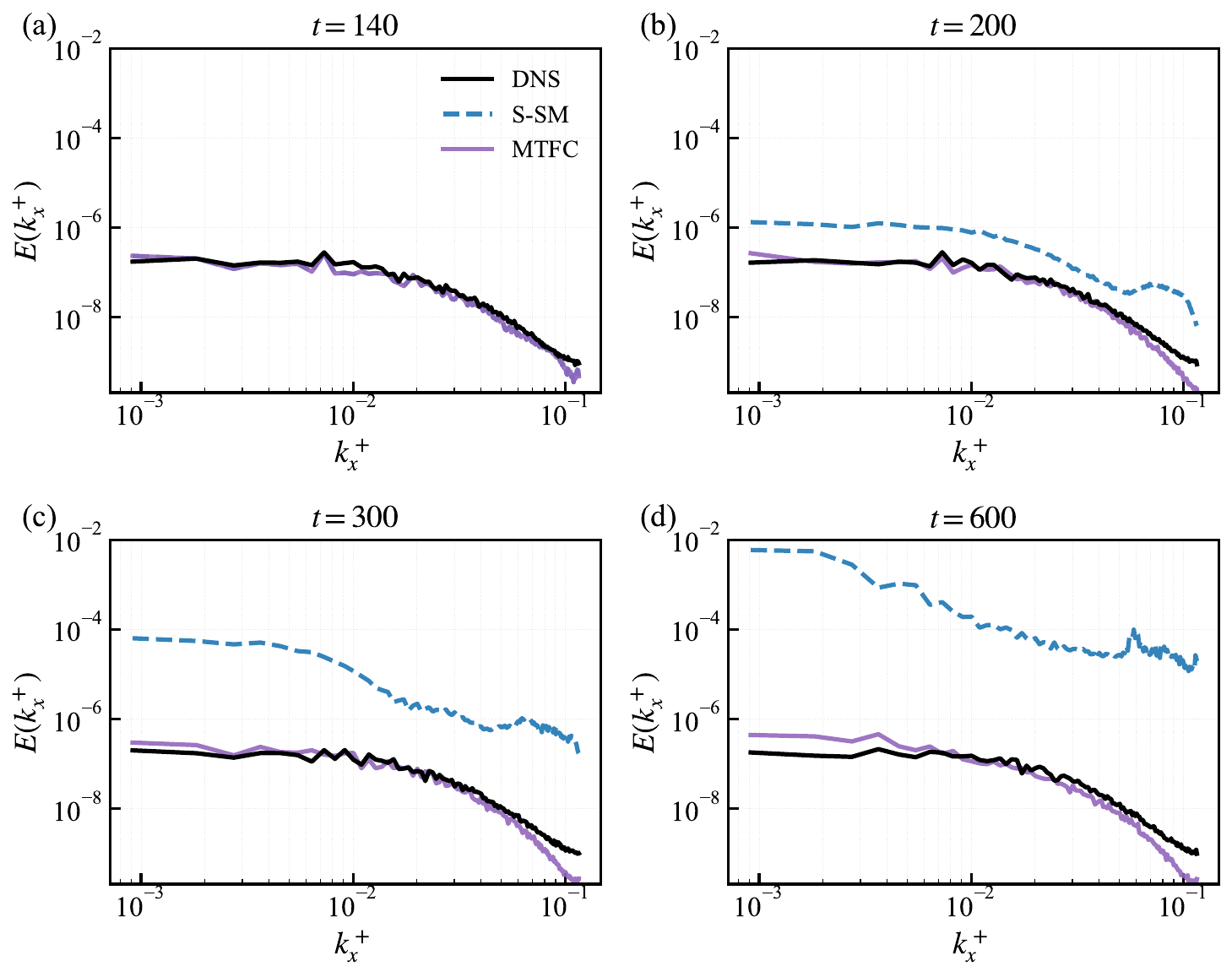}
	\caption{Streamwise energy spectra $E(k_x^+)$ during extended rollout at $t=140,\,200,\,300,\,600$ steps.}
	\label{fig:long_horizon_kx}
\end{figure}

\begin{figure}[htbp]
	\centering
	\includegraphics[width=\textwidth]{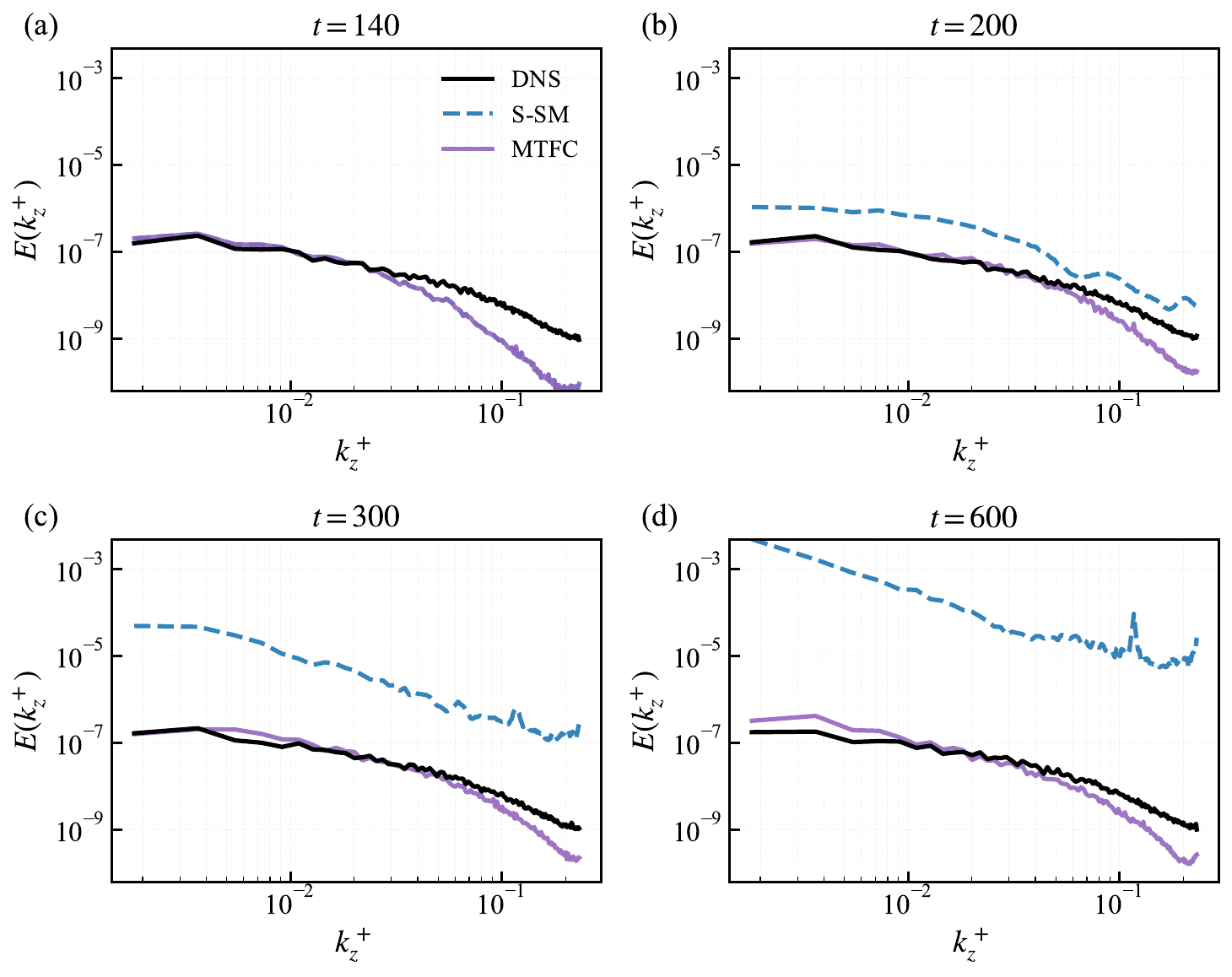}
	\caption{Spanwise energy spectra $E(k_z^+)$ during extended rollout at $t=140,\,200,\,300,\,600$ steps.}
	\label{fig:long_horizon_kz}
\end{figure}

Beyond maintaining low instantaneous error, reasonably accurate turbulence predictions must preserve key statistical properties over long-horizon predictions. Figs.~\ref{fig:long_horizon_kx} and~\ref{fig:long_horizon_kz} examine the streamwise and spanwise energy spectra $E(k_x^+)$ and $E(k_z^+)$ at four absolute rollout timestepss: $t=140$, 200, 300, and 600 steps. At $t=140$ (panel a), both S-SM and MTFC match DNS closely. Note that the rollout has not yet reached the first fusion point ($ffp=150$), so the two trajectories are still identical. At $t=200$ (panel b), S-SM has already begun to deviate from DNS\@. MTFC shows small deviations only at high $k_x^+$, corresponding to the small-scale structures of the turbulent fields, and remains well aligned with DNS over most of the spectrum. By $t=300$ (panel c) and $t=600$ (panel d), S-SM has departed from DNS by several orders of magnitude across the spectrum. MTFC continues to deviate only at the highest $k_x^+$ and preserves the DNS spectral shape elsewhere. The spanwise spectra $E(k_z^+)$ in Fig.~\ref{fig:long_horizon_kz} follow the same trends at all four times. Across both the streamwise and spanwise directions, MTFC reproduces similar spectra shapes to DNS over a much longer horizon than the S-SM alone.

\subsection{3D Turbulence Reconstruction}

The CTA-Swin-UNet and MTFC strategy demonstrated above provide stable, long-horizon predictions of velocity fields on a two-dimensional wall-parallel plane. However, many turbulence analyses and engineering applications require knowledge of the 3D flow volume. 
Training directly on 3D volumetric data would substantially increase the number of spatial tokens and memory demand, making end-to-end 3D autoregressive learning computationally expensive.
Instead, we apply the resolvent-based spectral linear stochastic estimation (SLSE) introduced in Section~\ref{sec:3drecon} to reconstruct the complete 3D velocity fields from the 2D planar predictions at each timestep, thereby recovering volumetric turbulent flow evolution without incurring the cost of end-to-end 3D network training.

\begin{figure}[htbp]
	\centering
	\includegraphics[width=\textwidth]{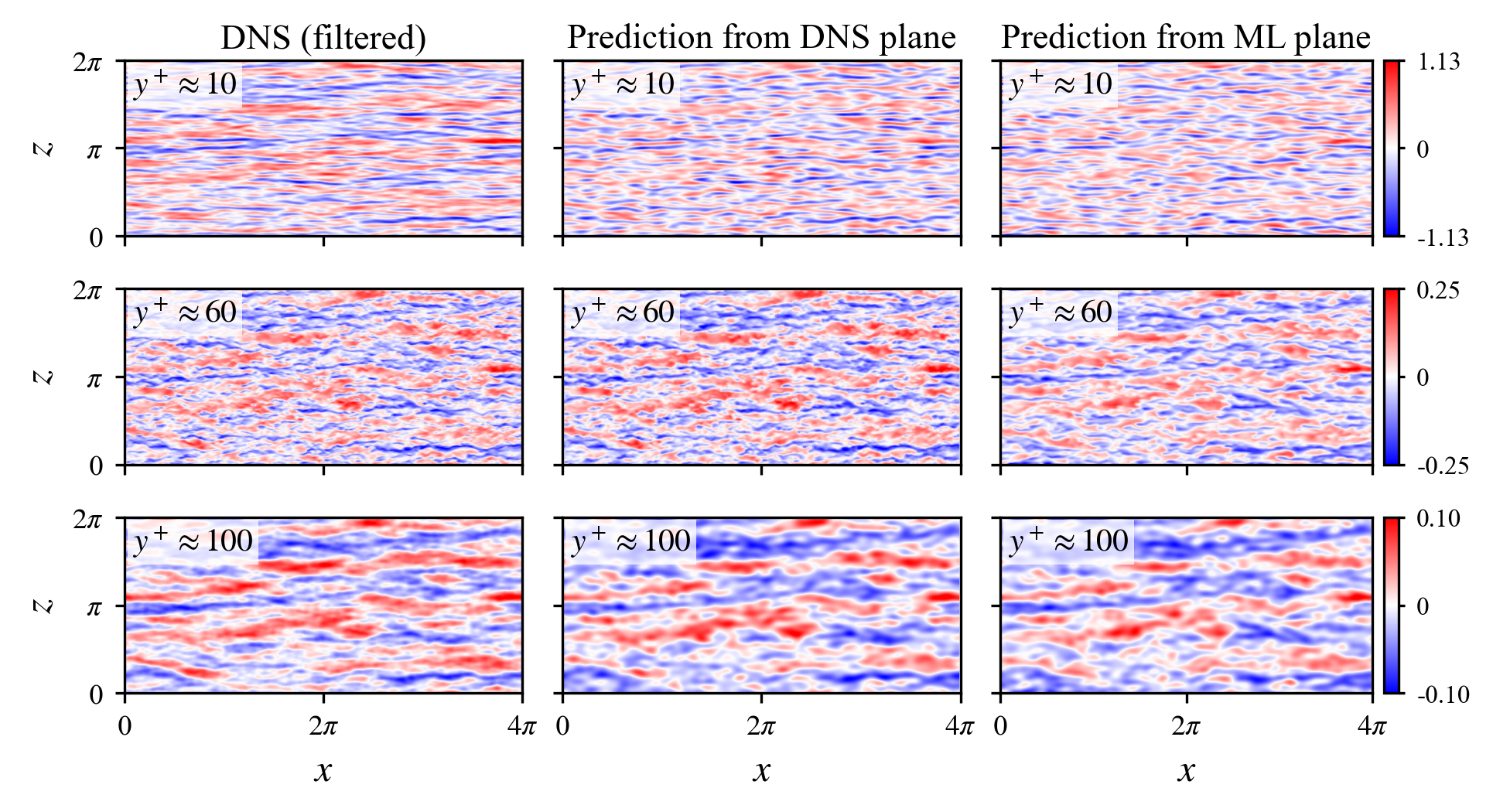}
	\caption{Instantaneous streamwise velocity on wall-parallel (XZ) planes at $t=200$ steps. Rows correspond to $y^+=10$, 60, and 100 from top to bottom. Left column (a, d, g): filtered DNS reference; center column (b, e, h): SLSE reconstruction using DNS measurements at the reference plane as input; right column (c, f, i): SLSE reconstruction using MTFC predictions at the reference plane as input.}
	\label{fig:recon_xz}
\end{figure}

The resolvent-based SLSE is applied frame-by-frame to the entire MTFC-predicted 2D time series, reconstructing the 3D velocity fields at every timestep of the extended rollout. For conciseness, we select results at $t=200$ steps as a representative instant within the MTFC stable regime. Figure~\ref{fig:recon_xz} presents a $3\times3$ comparison of the streamwise velocity at three wall-normal locations ($y^+=10$, 60, and 100). The left column (a, d, g) shows the filtered DNS reference. The center column (b, e, h) shows the resolvent-based SLSE reconstruction with the DNS fields at the reference plane $y^+=50$ as input. The right column (c, f, i) shows the resolvent-based SLSE reconstruction with the MTFC-predicted reference plane as input.
The comparison between the left and center columns isolates the error introduced by the resolvent-based SLSE itself. All three reconstructed planes recover the main streak structures of the corresponding DNS reference. The error at $y^+=60$ is markedly smaller than at $y^+=10$ and $y^+=100$. The latter two still capture the streak organization but show noticeable deviations in the velocity peaks and small-scale features. This is because the accuracy of the resolvent-based SLSE depends on the strength of the cross-plane spectral coherence. The plane at $y^+=60$ is the closest to the reference plane $y^+=50$, where the cross-plane correlation with the input is strongest.
The comparison between the right and center columns then shows that the difference between using DNS and using the MTFC prediction at the reference plane is small at all three wall-normal locations. The error in the MTFC-predicted reference plane is therefore not amplified when transmitted through the resolvent-based SLSE to other wall-normal planes, demonstrating the robustness of the proposed 3D prediction and reconstruction framework as a whole.

\begin{figure}[htbp]
	\centering
	\includegraphics[width=\textwidth]{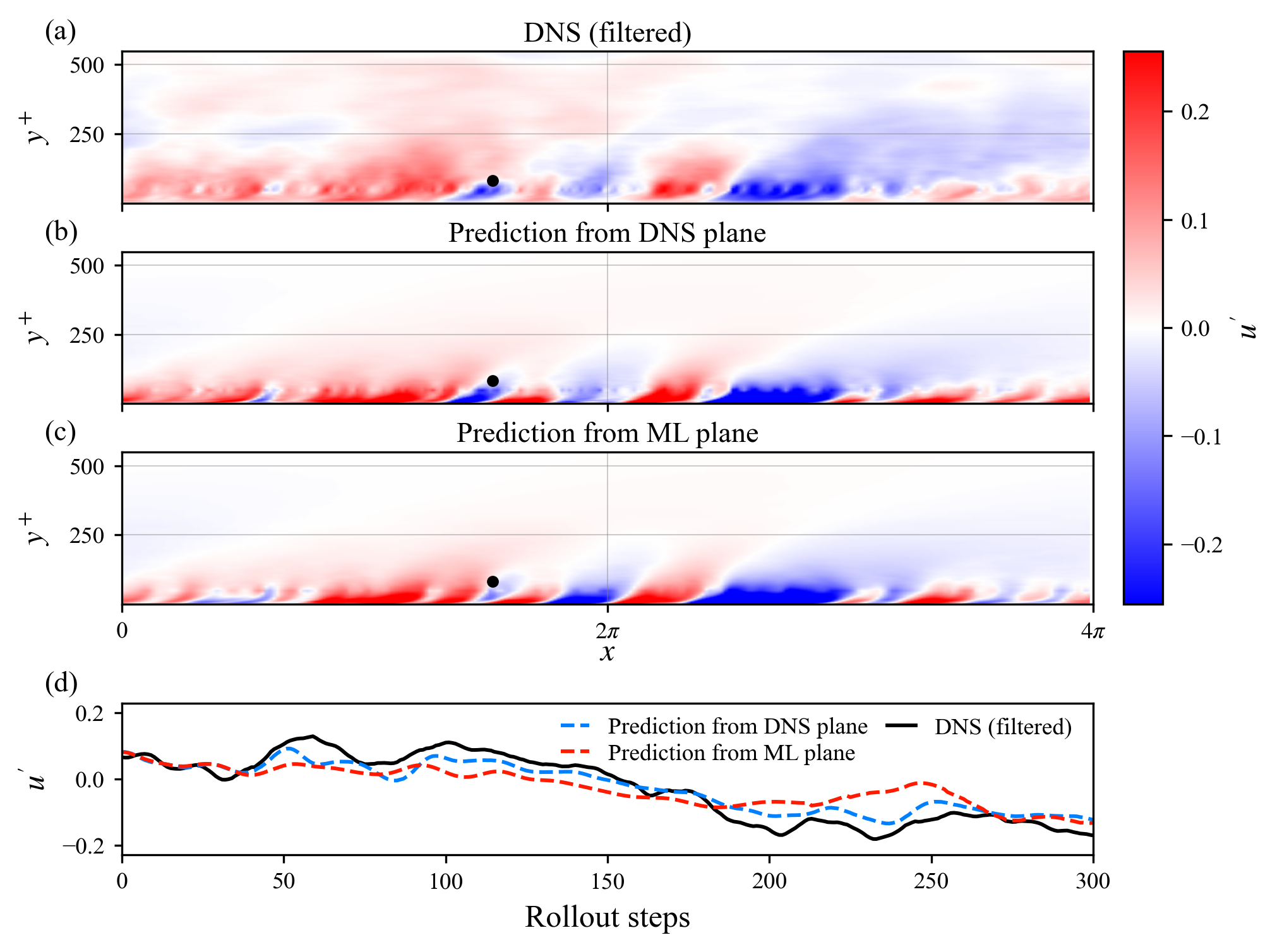}
	\caption{Streamwise velocity in the $(x,y)$ cross-section at spanwise position $z=0.5L_z$ and $t=200$ steps, together with probe time series. (a) Filtered DNS reference; (b) SLSE reconstruction using DNS measurements at the reference plane as input; (c) SLSE reconstruction using MTFC predictions at the reference plane as input. Black dots in panels (a)--(c) mark the probe sampling location. (d) Time evolution of the reconstructed streamwise velocity at a randomly selected probe location ($y^+=81.6$, $x=4.8$).}
	\label{fig:recon_xy}
\end{figure}

To complement the wall-parallel view, Fig.~\ref{fig:recon_xy} examines the reconstructed wall-normal velocity structure in an $(x,y)$ cross-section at fixed spanwise position $z=0.5L_z$. Panels (a), (b), and (c) show the filtered DNS reference, the SLSE reconstruction using DNS plane input, and the SLSE reconstruction using MTFC plane input at $t=200$, respectively.
Panels (b) and (c) reproduce the dominant wall-attached large-scale motions in the $(x,y)$ plane, manifested as inclined low- and high-momentum regions extending from the near-wall layer toward the channel core. The reconstructions preserve the wall-normal coherence and inclination of these energetic structures. 
The small-scale near-wall fluctuations are attenuated, as they are only weakly represented by the linearly coherent part of the SLSE transfer function.
To verify the temporal effectiveness of the reconstruction framework, Panel (d) shows the streamwise velocity time series at a randomly selected probe location ($y^+=81.6$, $x=4.8$, marked by black dots in panels a-c) over 300 rollout steps. The DNS reference (black solid), the SLSE reconstruction from DNS plane input (blue dashed), and the SLSE reconstruction from ML plane input (red dashed) track each other closely throughout the rollout. This confirms that the proposed 3D prediction and reconstruction framework remains accurate over the entire extended rollout.

\section{\label{sec:conclusions}Discussion and Conclusions}

We presented a hybrid machine-learning framework for long-horizon autoregressive prediction of 3D wall-bounded turbulence at $Re_\tau \approx 550$. The framework consists of three components, namely a CTA-Swin-UNet that performs one-step prediction on a wall-parallel plane at $y^+=50$, the MTFC strategy that periodically fuses predictions from a temporally coarser L-SM into the fine-scale rollout to suppress error accumulation, and a resolvent-based SLSE that reconstructs the corresponding 3D flow fields from the predicted planar time series.

In single-step prediction, the CTA-Swin-UNet attains the lowest test-set errors among the four compared models across all three velocity components. Under autoregressive rollout, it maintains spatial and spectral fidelity for approximately 150 rollout steps, whereas the plain Swin-UNet, LSTM, and FNO baselines lose predictive capability within roughly 20 to 50 rollout steps. Removing the CTA module degrades single-step accuracy only moderately but accelerates divergence under rollout, indicating that the channel-time-attention mechanism mainly stabilizes error propagation across recursive predictions rather than improving single-step accuracy. With MTFC, the stable prediction window extends beyond 300 rollout steps, and the streamwise and spanwise energy spectra remain aligned with DNS up to $t=600$ rollout steps, while the S-SM alone develops spectral inflation of two to three orders of magnitude over the same horizon. Applied frame-by-frame to the MTFC predictions, the resolvent-based SLSE recovers the streak organization on wall-parallel planes at $y^+=10$, $60$, and $100$, and the inclined high- and low-momentum regions in cross-sectional $(x,y)$ views. The reconstruction error introduced by feeding MTFC predictions in place of DNS fields at the reference plane remains small at all three wall-normal locations.

The limitations of this work should be noted. 
The present results are obtained at a single Reynolds number and a single reference-plane height, and the generalization across different $Re_\tau$ and reference-plane locations remains to be verified. 
In addition, the linear nature of the SLSE transfer function attenuates small-scale fluctuations with weak cross-plane spectral coherence, thereby limiting the recovery of fine-scale structures away from the reference plane. 
Future work will explore nonlinear extensions of the SLSE for improved small-scale reconstruction, further ablation studies on the CTA ordering and the MTFC fusion weight $\alpha$, and validation using higher-Reynolds-number datasets.

\nocite{*}

\bibliographystyle{aipnum4-1}
\bibliography{refs}

\appendix

\section{\label{sec:hyperparams}Hyperparameter Configuration}

This appendix lists the network architecture hyperparameters for all models evaluated in this work, namely the proposed CTA-Swin-UNet (S-SM), the ablation counterpart Swin-UNet-NoAttn (identical structure to S-SM but with the channel-time-attention module disabled), the L-SM used in MTFC (trained with a temporal sampling stride 10 times that of the S-SM to capture slower large-scale dynamics), LSTM, and FNO. 

\begin{table}[htbp]
\caption{\label{tab:hyper-lstm}Architecture hyperparameters of the LSTM baseline.}
\begin{ruledtabular}
\begin{tabular}{lclc}
\textbf{Hyperparameter} & \textbf{Value} & \textbf{Hyperparameter} & \textbf{Value} \\
\hline
Network type & ConvLSTM & Hidden dim.\ sequence & [16,16,32,32,64] \\
No.\ of ConvLSTM layers & 3 & Conv.\ kernel size & [3,3,3] \\
Bias & Yes & Input frames $L$ & 5 \\
Input components $C$ & 3 & Spatial resolution & $256\times256$ \\
\end{tabular}
\end{ruledtabular}
\end{table}

\begin{table}[htbp]
\caption{\label{tab:hyper-fno}Architecture hyperparameters of the FNO baseline.}
\begin{ruledtabular}
\begin{tabular}{lclc}
\textbf{Hyperparameter} & \textbf{Value} & \textbf{Hyperparameter} & \textbf{Value} \\
\hline
Hidden channels & 128 & Fourier modes & $64\times64$ \\
No.\ of spectral layers & 4 & Lifting channels & 128 \\
Projection channels & 128 & Input components $C$ & 3 \\
\end{tabular}
\end{ruledtabular}
\end{table}

\begin{table}[htbp]
\caption{\label{tab:hyper-ssm}Architecture hyperparameters of the CTA-Swin-UNet (S-SM).}
\begin{ruledtabular}
\begin{tabular}{lclc}
\textbf{Hyperparameter} & \textbf{Value} & \textbf{Hyperparameter} & \textbf{Value} \\
\hline
Embedding dim. & 192 & Patch size & $4\times4$ \\
Stage depths & [2,4,4,6,4,4,2] & No.\ of attn.\ heads & 32 \\
Window size & $8\times8$ & MLP expansion ratio & 4.0 \\
Dropout rate & 0.1 & Attn.\ dropout rate & 0.1 \\
Drop-path rate & 0.1 & QKV bias & Yes \\
Temporal stride & 1 & Spatial resolution & $256\times256$ \\
\end{tabular}
\end{ruledtabular}
\end{table}

\begin{table}[htbp]
\caption{\label{tab:hyper-lsm}Architecture hyperparameters of the CTA-Swin-UNet (L-SM).}
\begin{ruledtabular}
\begin{tabular}{lclc}
\textbf{Hyperparameter} & \textbf{Value} & \textbf{Hyperparameter} & \textbf{Value} \\
\hline
Embedding dim. & 160 & Patch size & $4\times4$ \\
Stage depths & [2,2,4,6,4,2,2] & No.\ of attn.\ heads & 32 \\
Window size & $8\times8$ & MLP expansion ratio & 4.0 \\
Dropout rate & 0.1 & Attn.\ dropout rate & 0.1 \\
Drop-path rate & 0.1 & QKV bias & Yes \\
Temporal stride & $10\times$ S-SM & Spatial resolution & $256\times256$ \\
\end{tabular}
\end{ruledtabular}
\end{table}

\end{document}